\documentclass[english,tightenlines,eqsecnum,amsmath,amssymb,aps,prd,superscriptaddress,nofootinbib,showpacs,floats]{article} 
\usepackage{amsmath,amsfonts,amssymb,multicol}
\usepackage{leftidx}
\usepackage{tensor}
\usepackage{graphicx,caption,subcaption}
\usepackage[usenames]{xcolor}
\usepackage{epstopdf}
\usepackage{graphicx}
\usepackage{lipsum}
\usepackage{euscript,amssymb}
\usepackage{epsfig}
\usepackage{fancyhdr}
\usepackage{esint}
\usepackage{color}

{\definecolor{RED}{rgb}{1,0,0}
\definecolor{GREEN}{rgb}{0,1,0}
\definecolor{BLUE}{rgb}{0,0,1}

\def\be{\begin{equation}}
\def\ee{\end{equation}}
\def\bea{\begin{eqnarray}}
\def\eea{\end{eqnarray}}
\usepackage{amsfonts}
\usepackage{amsmath}
\usepackage{amssymb}
\usepackage{fancyhdr}
\def\nn{\nonumber\\}
\def\be{\begin{equation}}
\def\ee{\end{equation}}
\def\bea{\begin{eqnarray}}
\def\eea{\end{eqnarray}}

\def\bwt{\begin{widetext}}
\def\ewt{\end{widetext}}

\usepackage{esint}
\usepackage[unicode=true, pdfusetitle,
 bookmarks=true,bookmarksnumbered=false,bookmarksopen=false,
 breaklinks=false,pdfborder={0 0 1},backref=false,colorlinks=false]
 {hyperref}
\def\T{T}

\newcommand{\f}[2]{\frac{#1}{#2}}
\begin{document}
\title{Traversable wormholes in Einsteinian cubic gravity}
\author{\hspace{-1cm}Mohammad Reza Mehdizadeh$^{1\!\!}$ \footnote{mehdizadeh.mr@uk.ac.ir}\,\, and\, Amir Hadi Ziaie$^2$\footnote{ah.ziaie@maragheh.ac.ir}
	\\\\\hspace{-1cm}$\leftidx{^1}{{\rm Department~of~ Physics,~ Shahid~ Bahonar~ University, P.O.~ Box~ 76175, Kerman, Iran}}$ 
	\\\\
	\hspace{-1cm}{$^2$Research~Institute~for~Astronomy~and~Astrophysics~of~ Maragha~(RIAAM),~University~of~Maragheh,}\\{P.~O.~Box~55136-553~Maragheh,~Iran}}
\date{\today}
\maketitle
\begin{abstract}
In the present work we investigate  wormhole configurations described by a constant redshift function in Einstein-Cubic gravity ({{\sf ECG}}). We derive analytical wormhole geometries by assuming a particular equation of state ({{\sf EoS}}) and investigate the possibility that these solutions satisfy the standard energy conditions. We introduce exact asymptotically flat and anti-de Sitter (AdS) spacetimes that admit traversable wormholes. These solutions are obtained by imposing suitable values for the parameters of the theory so that the resulted geometries satisfy the weak energy condition ({\sf WEC}) in the vicinity of the throat, due to the presence of higher-order curvature terms. Moreover, we find that AdS solutions satisfy the {\sf WEC} throughout the spacetime. A description of the geodesic motion of timelike and null particles is presented for the obtained wormhole solutions. Also, using gravitational lensing effects, observational features of the wormhole structure are discussed.
\end{abstract}
\section{Introduction}
A wormhole is a hypothetical topological feature of the
spacetime which creates shortcuts between two distinct spacetimes. The
concept of \lq\lq{}wormhole\rq\rq{} for these structures was invented for the first time in $1957$ within the seminal papers of Misner and Wheeler \cite{misner-wheeler} and Wheeler \cite{Wheelerworm}, in order to provide a mechanism for having \lq\lq{}charge without charge\rq\rq{}. The study of Lorentzian wormholes in the context of General Relativity {\sf GR} dates back to the fundamental works of Morris and Thorne in 1988 \cite{mt} where, introducing a static spherically symmetric line element, they showed that, exact solutions representing wormhole geometries could be found by solving the Einstein field equation. According to {\sf GR}, the fundamental faring-out condition of throat causes to the violation of null energy condition ({\sf NEC}) and thus, traversable wormholes are only possible if the so called \lq\lq{}{\textrm exotic matter}\rq\rq{} \cite{khu} exists at their throat, which involves an energy-momentum tensor ({\sf EMT}) violating the {\sf NEC}. This condition is in turn a part of the {\sf WEC} that the physical meaning of which is that the energy density be non-negative in any reference frame. In this respect, traversable wormhole geometries have been obtained using the exotic matter distribution, e.g., with the help of phantom energy distribution \cite{phantworm}. This type of matter, though exotic in the laboratory context, is of observational interest in cosmological scenarios \cite{phan-dark-sce}. Phantom energy possesses peculiar attributes, namely, a divergent cosmic energy density in a finite time\cite {phant1}, prediction of existence of a new long range force \cite{phant2}, and the appearance of a negative entropy and negative temperature~\cite{phant3}.
\par
One of the most significant issues in wormhole geometries is the fulfillment of standard energy conditions. In this regard, many attempts have been devoted in the literature in order to find some realistic material sources that support the
wormhole configuration or minimize the employment of exotic matter. Work along this line have been done to investigate the construction of thin-shell, dynamical, and rotating wormholes~\cite{thindynrot}. However, we concentrate, in the herein model, on modified theories of gravity where the effective scenario could provide static spherically symmetric solutions representing traversable wormholes without resorting to exotic matter source. In this context, it has been shown that the presence of higher order terms in curvature would allow for constructing thin-shell wormholes supported by ordinary matter \cite{thi}. Moreover, wormholes in modified gravity involving higher-order curvature invariants can satisfy the energy conditions at least at the throat \cite{glo1} and throughout the spacetime \cite{mkl}. The study of wormhole solutions in the framework of modified gravity has become a main focus of interest in modern cosmology. This topic has attracted much attention in recent years and a large amount of work has been devoted to this issue among which we quote: spacetimes admitting wormhole solutions in Brans-Dicke theory \cite{bd}, $f(R)$ gravity \cite{fr}, Born-Infeld theory~\cite{bf}, Einstein-Gauss-Bonnet theory~\cite{gmfl}, Kaluza-Klein gravity~\cite{kl}, Rastall gravity~\cite{rastallworm}, scalar-tensor gravity~\cite{kash} wormhole solutions in the presence of a
cosmological constant~\cite{anac}, non-commutative geometry~\cite{Wormnonc} and other modified gravity theories~\cite{modgrworm}.
\par
Over recent years there has been a remarkable interest in the subject of higher curvature gravity, that the main focus of much has been motivated by the wish to account for quantum phenomena within gravitational field. In~\cite{ste1}, it was shown that including higher curvature terms within the Einstein-Hilbert action can provide a setting towards a renormalizable theory of gravity. It is generally expected that higher order terms show up themselves, e.g., within the renormalization of quantum field theory in curved spacetime~\cite{QFTCST}, or in the construction of low-energy effective action of string theory~\cite{LEEAST} such as Lovelock theory~\cite{LLOVE}. Moreover, in order that the gravitational action be free of ghost terms, the quadratic curvature corrections to the Einstein-Hilbert action should be proportional to the Gauss-Bonnet term~\cite{GBTERM}. Also, gravity theories with higher curvature corrections have attracted a great deal of interest in holography~\cite{hol1} and considered in the context of cosmology~\cite{cos1}.
\par
A particular proper feature of a specified higher-order gravity is its linearized spectrum, that is, the set of physical degrees of freedom propagating through metric perturbations in vacuum. For instance, in the framework of holography, useful information about the corresponding holographic {\sf CFT} stress-energy tensors can be extracted by the virtue of linearized equations of a given higher-order gravity as these equations are dual to the metric perturbation, see e.g.,~\cite{cfteqs}. There are some higher-order gravity theories such as quasi-topological gravity~\cite{qttheory} and $f({\rm Lovelock})$~\cite{flovel} which are equivalent to {\sf GR} at the linearized level in the vacuum, i.e., the only physical mode propagated by the metric perturbations is a transverse and massless graviton. However, the inconvenient characteristic of most of higher-order gravity theories is the dependency of the couplings of the different curvature invariants on spacetime dimension and in fact, these theories are different gravity theories in different dimensions. Nevertheless, the only known theories that their couplings are independent of spacetime dimension and share spectrum with {\sf GR} are
Lovelock theories~\cite{LLOVE}. In this regard, it is shown that, up to cubic order corrections in curvature, there is only one additional theory which satisfies this criterion. Furthermore, this theory is non-trivial in four dimensions, unlike the quadratic and cubic Lovelock theories. Recently, a new model of higher order gravity has been presented in~\cite{con1} to which, when quadratic and cubic Lovelock terms are added, is the unique cubic model for gravitational interaction that shares its graviton spectrum with {\sf GR}. Moreover it is shown that this theory has dimension-independent coupling constants and is free of massive gravitons in general dimensions~\cite{ghostfreecubic}. In the context of this theory, which is known as Einsteinian cubic gravity ({\sf ECG}), black hole solutions in four-dimensions~\cite{cub4} and in higher dimensional spacetimes have been studied~\cite{rub1}. The asymptotically-AdS black brane solutions of {\sf ECG} have also been considered in \cite{high2}.

\par
Motivated by the above considerations we search for static spherically symmetric solutions representing wormhole configurations in {\sf ECG} and study the effects of higher curvature terms within the wormhole structure. This paper is organized as follows: In section~\ref{ECG}, we give a brief review on gravitational field equations in {\sf ECG}, and proceed with analyzing the the energy conditions using a general form for spacetime metric of a worm hole. In section~\ref{WHS}, we take
an {\sf EoS} for the radial and tangential pressures and search for exact wormhole solutions along with checking the satisfaction of energy conditions. Our conclusion is drawn in section~\ref{concluding}.
\section{Action and Field equations}\label{ECG}
The action of pure Einsteinian Cubic gravity in four dimensions is given by \cite{con1}
\bea {\sf S}&=&\int  d^4x\sqrt{-g} \left({\mathcal L}_g +{\mathcal L}_m\right),~~~~{\mathcal L}_g=\frac{1}{2\kappa} \left[-2 \Lambda + {\sf R} \right] + \kappa \lambda {\cal P}, \label{action}
\eea
where ${\mathcal L}_m$ and ${\mathcal L}_g$ are the material and gravitational parts of the action, respectively, $\kappa=8\pi G$ is the gravitational coupling constant and $\lambda$ is a dimensionless coupling constant. Also, {\sf R} is the Ricci scalar and the new Einsteinian cubic gravity term  ${\cal P}$ contributes to the action as cubic curvature terms in four dimensions, defined as 
\begin{align}
{\cal P} = 12 \tensor{\sf R}{_a ^c _b ^d} \tensor{\sf R}{_c ^m _d ^n}\tensor{\sf R}{_m ^a _n ^b} + {\sf R}_{\,\,\,\,\,{\sf cd}}^{{\sf ab}} {\sf R}_{\,\,\,\,\,\,{\sf ab}}^{{\sf mn}}{\sf R}_{\,\,\,\,\,{\sf mn}}^{{\sf cd}} - 12 {\sf R}_{{\sf ab c d}}{\sf R}^{{\sf a} {\sf c}}{\sf R}^{{\sf bd}} + 8 {\sf R}_{\,\,{\sf a}}^{\sf b} {\sf R}_{\,\,{\sf b}} ^{\sf c} {\sf R}_{\,\,{\sf c}} ^{\sf a},
\end{align}
where ${\sf R}_{{\sf ab c d}}$ is the Riemann curvature tensor constructed out of Christoffel symbols. The field equations of {\sf ECG} can then be written as~\cite{con1,ghostfreecubic}
\begin{equation}
{\sf P}_{{\sf a cde}}{\sf R}_{{\sf b}}{}^{{\sf cde}} - \frac{1}{2}{\sf g}_{{\sf ab}} {\cal L} - 2 \nabla^{\sf c} \nabla^{\sf d} {\sf P}_{{\sf acdb}} =\frac{1}{2}{\sf T}_{{\sf a b}}\, ,
\label{fieldequations}
\end{equation}
where
\begin{align}\label{P_thing}
{\sf P_{abcd}} =& \,  \frac{\partial {\cal L}}{\partial {\sf R^{abcd}}} \,, \nn
=& \frac{1}{2 \kappa} {\sf g}_{{\sf a}[{\sf c}}{\sf g}_{{\sf b}]{\sf d}} + 6 \kappa \lambda \big[  \,  {\sf R _{ad} R _{bc}} -  {\sf R_{ac} R _{bd}} +  {\sf g_{bd}} {\sf R _{a}{}^{e}} \ 
{\sf R _{ce}} -  {\sf g_{ad} R _{b}{}^{e} R_{ce}}  -  {\sf g_{bc} R _{a}{}^{e} R_{de}}  \nn
&+  {\sf g_{ac} R _{b}{}^{e} R_{de}}
-  {\sf g_{bd} R ^{ef} R_{aecf}} +  {\sf g_{bc} R ^{ef} R _{aedf}} + \
{\sf g_{ad} R ^{ef} R _{becf}} - 3 {\sf R_{a}{}^{e}{}_{d}{}^{f} R _{becf}} \nn
&  - {\sf g_{ac} R ^{ef} R _{bedf}} + 3 {\sf R_{a}{}^{e}{}_{c}{}^{f} R _{bedf}} + \tfrac{1}{2} {\sf R_{ab}{}^{ef} R _{cdef}} \big] \, ,
\end{align}
and ${\sf T_{a b}}$ is the {\sf EMT} of matter fields.
In this work, we are interested in wormhole solutions in four-dimensions. Therefore, we proceed with employing a general static spherically symmetric line element which represents a wormhole and is given by
\begin{eqnarray}\label{WHmetric}
ds^2= -e^{2\psi(r)}dt^2+
\frac{dr^2}{1-\frac{b(r)}{r}}+r^2( d\theta^2 + \sin^2 \theta d\phi^2 ),
\end{eqnarray}
where $\psi(r)$ being the redshift function and $b(r)$ is the wormhole's shape function. The shape function must satisfy the flare-out condition at the throat, i.e., we must have $b^{\prime}(r_0)<1$ and $b(r)<r$ for $r>r_0$ in the whole spacetime, where $r_0$ is the throat radius. In the present work, we consider $\psi(r)=0$ in order to ensure the absence of horizons and singularities throughout the spacetime. The {\sf EMT} of matter is given by the following diagonal form
\begin{equation}
{\sf T}_{\,\,\,{\sf b} }^{{\sf a} }=\mathrm{diag} \left[-\rho\left(r\right),p_{r}\left(r\right),p_{t}\left(r\right),p_{t}\left(r\right)\right] \,,
\label{def:SET}
\end{equation}
where $\rho(r)$ is the energy density and $p_{r}(r)$ and $p_{t}(r)$ are the radial and tangential pressures, respectively. Using field equation (\ref{fieldequations}) along with taking into account the metric (\ref{WHmetric}) we get
\begin{eqnarray}
\rho(r)& =&\frac{b^{\prime}-\Lambda r^2}{\kappa r^2}+\frac{\kappa\lambda}{r^9}\Bigg\{24(r-b)r^3\left(b^{\prime \prime \prime }(2b-b^{\prime }r)-rb^{\prime \prime2}\right)  \notag \\	&&-12r^2b^{\prime \prime}\left(-r^2b^{\prime2}+rb^{\prime}\left(21b-18r\right)-28b^2+26rb\right)\notag \\	&&+6(b^{\prime}r-b)\left(5r^2b^{\prime2}+b^{\prime}\left(58r^2-72rb\right)+90rb+99b^2\right)\Bigg\}
, \label{rho}
\end{eqnarray}

\begin{eqnarray}
p_{r}(r) =\frac{r^3\Lambda-b(r)}{\kappa r^3},  \label{tau}
\end{eqnarray}
	\begin{eqnarray}
p_{t}(r)=\frac{b(r)-rb^{\prime}(r)+2r^3\Lambda}{2\kappa r^3}, \label{pr}
\end{eqnarray}
where a prime denotes derivative with respect to the radial coordinate $r$. It is notable that the radial and tangential pressure profiles are exactly as in the {\sf GR} case and the higher order curvature terms within the action and field equations (\ref{fieldequations}) of the background theory contribute only to the energy density. This is due to the assumption for vanishing redshift function and for $\psi(r)\neq0$, one gets more complicated expressions for pressure profiles. In view of the local energy conditions, we examine the {\sf WEC}, which asserts that ${\sf T}_{{\sf ab}}{\sf U}^{{\sf a}}{\sf U}^{{\sf b}}\geq 0$ where ${\sf U}^{{\sf a}}$ is a timelike vector field. For the diagonal {\sf EMT} (\ref{def:SET}), the {\sf WEC} implies $\rho \geq 0$, $\rho +p_{r}\geq 0\ $and $\rho +p_{t}\geq 0$. Note that the {\sf WEC} implies the {\sf NEC}. Using equations (\ref{rho})-(\ref{pr}), we arrive at the following relations 
 \begin{eqnarray}
\rho +p_{r}&=& \frac{\kappa\lambda}{r^9}\Bigg\{24(r-b)r^3\left(b^{\prime \prime \prime }(2b-b^{\prime }r)-rb^{\prime \prime2}\right)  \notag \\	&&-12r^2b^{\prime \prime}\left(-r^2b^{\prime2}+rb^{\prime}\left(21b-18r\right)-28b^2+26rb\right)\notag \\	&&+6(b^{\prime}r-b)\left(5r^2b^{\prime2}+b^{\prime}\left(58r^2-72rb\right)+90rb+99b^2\right)\Bigg\}\notag \nn&&+\frac{(b^{\prime}r-b)}{\kappa r^3},
\label{EGBNEC}  
\end{eqnarray}%
\begin{eqnarray}
\rho +p_{t}&=&\frac{\kappa\lambda}{r^9}\Bigg\{24(r-b)r^3\left(b^{\prime \prime \prime }(2b-b^{\prime }r)-rb^{\prime \prime2}\right)  \notag \\	&&-12r^2b^{\prime \prime}\left(-r^2b^{\prime2}+rb^{\prime}\left(21b-18r\right)-28b^2+26rb\right)\notag \\	&&+6(b^{\prime}r-b)\left(5r^2b^{\prime2}+b^{\prime}\left(58r^2-72rb\right)+90rb+99b^2\right)\Bigg\}+\frac{b^{\prime}r+b}{2\kappa r^3}.
\end{eqnarray}
One can easily show that for $\lambda=0$  the {\sf NEC}, and consequently the {\sf WEC} are violated at the throat ($\rho +p_{r}<0$), due to the
flaring-out condition~\cite{misner-wheeler,Wheelerworm}. Note that at the throat, one verifies
\begin{eqnarray}
\left( \rho + p_r \right)\big|_{r=r_0}=- \frac{1}{\kappa r_0^2}\left(
1-b^{\prime }_0 \right)+\frac{6\lambda \kappa}{r_0^6}\left(1-b^{\prime }_0 \right)\left[5({b^{\prime }_0}^2+1)-2b^{\prime }_0(b^{\prime \prime}_0r_0+7)+4b^{\prime \prime}_0 r_0+4\right],
\label{WECthroat}
\end{eqnarray}
which shows that for $\lambda=0$ the {\sf NEC}, and consequently the {\sf WEC}, are violated at the throat. In order to impose $ \rho +p_{r}>0$ in {\sf ECG}, it is now possible to find  an adequate range for the model parameters such that the
{\sf NEC} is satisfied at the throat. In the following section, we
search for exact wormhole solutions in {\sf ECG} and investigate in detail the conditions under which the {\sf WEC} is respected.
\section{Wormhole Geometries}\label{WHS}
In this section we present exact wormhole solutions in the context of {\sf ECG}. Here, we have a system of differential equations (\ref{rho})-(\ref{pr}) with four unknown functions $\rho(r)$, $p_r(r)$, $p_t(r)$ and $b(r)$. In order to determine spacetimes admitting wormhole structures, we must adopt a strategy for specifying the shape function  $b(r)$. We may also consider a pre-determined form for the
functionality of $b(r)$ and consequently obtain the {\sf EMT} components. Also, it is common to consider a specific form for the equation of state ({\sf EoS}), which provides a relation between the {\sf EMT} components, namely, a linear {\sf EoS} ~\cite{linea1}, the traceless~\cite{trc1} and {\sf EoS} in general form~\cite{anch1}. However a linear {\sf EoS} involving energy density leads to a complicated differential equation. This is due to the presence of third order derivative for $b(r)$. Thus, we proceed with a less complicated differential equation for the shape function and consider an {\sf EoS} relating the tangential and radial  pressures, so that the energy density is determined through Eq. (\ref{rho}). Here, we consider an {\sf EoS} of the form $p_t=\alpha p_r$~\cite{frh}, whereby using Eqs. (\ref{tau})-(\ref{pr}) we obtain the following differential equation for the shape function
\begin{eqnarray}
b^{\prime}r-b(2\alpha+1)+2\Lambda(\alpha-1)r^3=0, \label{solution1}
\end{eqnarray}
for which the solution reads
\begin{eqnarray}
b(r) = \Lambda r^3 + {\sf C}_0 r^{2\alpha +1},\label{sol1}
\end{eqnarray}
where the integration constant ${\sf C}_0$ can be determined from the condition $b(r_0)=r_0$ at the throat and is given by 
\begin{eqnarray}
{\sf C}_0 = (1-\Lambda r_0^2)r_0^{-2\alpha}.
\end{eqnarray}
The flare-out condition at the throat results in the following inequality 
\begin{eqnarray}
b^{\prime}(r_0)-1= 2 r_0^2(1-\alpha)\Lambda+2\alpha<0, \label{bpr1}
\end{eqnarray}
and the energy density for this solution is found as
\begin{eqnarray}
\rho(r)= [\xi_3\Lambda^3+\xi_2\Lambda^2+\xi_1\Lambda+\xi_0]\kappa \lambda+\f{\Lambda}{\kappa}\left[2-(2\alpha+1)\left(\frac{r}{r_0}\right)^{2(\alpha-1)}\right]+\f{(2\alpha+1)}{\kappa r_0}\left(\frac{r}{r_0}\right)^{(2\alpha-1)},\label{rho11}
\end{eqnarray}
where 
\begin{eqnarray}
\xi_3 &=& \Bigg\{48(16\alpha^3-6\alpha^2+3\alpha-1)\left(\frac{r}{r_0}\right)^{4(\alpha-1)}+48\alpha\left(20\alpha^2-17\alpha+3\right)\left(\frac{r}{r_0}\right)^{6(\alpha-1)}\notag \\&& +96(2\alpha+1)\left(\frac{r}{r_0}\right)^{2(\alpha-1)} \Bigg\}(1-\alpha),
\end{eqnarray}

\begin{eqnarray}
\xi_2 &=& \Bigg\{ 96(16\alpha^3-6\alpha^2+3\alpha-1)\left(\frac{r}{r_0}\right)^{4(\alpha-1)}+\f{48}{r_0^2}(1-4\alpha^2) \left(\frac{r}{r_0}\right)^{2(\alpha-2)} -\f{96\alpha}{r_0^{2}}(2\alpha+1)\left(\frac{r}{r_0}\right)^{2(\alpha-1)} \notag \\&& -\f{48\alpha}{r_0^{2}}(5\alpha-3)(4\alpha-1)\left(\frac{r}{r_0}\right)^{6(\alpha-1)}+\f{48\alpha}{r_0^{2}}(4\alpha-1)(4\alpha-3)\left(\frac{r}{r_0}\right)^{2(2\alpha-3)}\Bigg\}(1-\alpha),
\end{eqnarray}

\begin{eqnarray}
\xi_1 &=& \Bigg\{\f{48}{r_0^{4}}(16\alpha^3-6\alpha^2+3\alpha-1)\left(\frac{r}{r_0}\right)^{2(2\alpha-2)}-\f{48\alpha}{r_0^{4}}(5\alpha-3)(4\alpha-1)\left(\frac{r}{r_0}\right)^{6(\alpha-1)}\notag \\&&+\f{96\alpha}{r_0^{4}}(4\alpha-1)(4\alpha-3)\left(\frac{r}{r_0}\right)^{2(2\alpha-3)}+\f{48}{r_0^{4}}(1-4\alpha^2)\left(\frac{r}{r_0}\right)^{2(\alpha-2)}\Bigg\}(\alpha-1),
\end{eqnarray}

\begin{eqnarray}
\xi_0=48\alpha(\alpha-1)(4\alpha-1)\Bigg\{(5\alpha-3)\left(\frac{r}{r_0}\right)^{6(\alpha-1)}-(4\alpha-3)\left(\frac{r}{r_0}\right)^{(4\alpha-6)}\Bigg\}r_0^{-6}.
\end{eqnarray}
We also get the following expressions for radial and tangential pressure profiles, as
\begin{eqnarray}
p_{r}&=&\frac{\left(\Lambda r_0^2-1\right)}{\kappa r_0^{2}}\left(\frac{r}{r_0}\right)^{2(\alpha-1)},
\label{pe1} 
\end{eqnarray}
\begin{eqnarray}
p_{t}&=& \frac{\alpha \left(\Lambda r_0^2-1\right)}{\kappa r_0^{2}} \left(\frac{r}{r_0}\right)^{2(\alpha-1)}.
\label{te1} 
\end{eqnarray}
From the expressions (\ref{rho11}), (\ref{pe1}) and (\ref{te1}) we can find the energy conditions at the throat as
\begin{eqnarray}
\rho(r_0)= \frac{(1-2\alpha)\Lambda r_0^2+(1+2\alpha)}{\kappa r_0^2}+\kappa \lambda \xi, \label{rho00}
\end{eqnarray}

\begin{eqnarray}
\rho(r_0)+p_r(r_0)=\frac{2(1-\alpha)\Lambda r_0^2+2\alpha }{\kappa r_0^2}+\kappa \lambda \xi, \label{pr00}
\end{eqnarray}
\begin{eqnarray}
\rho(r_0)+p_t(r_0)=\frac{(1-\alpha)\Lambda r_0^2+(1+\alpha)}{\kappa r_0^2}+\kappa \lambda \xi, \label{pt00}
\end{eqnarray}
where
\begin{eqnarray}
\xi& =& -48(\alpha-1)^3(4\alpha+1)\Lambda^3+\frac{48(12\alpha^2-3\alpha-1)(\alpha-1)^2}{r_0^2}\Lambda^2\notag \\&&-\frac{144\alpha^2 (4\alpha^2-7\alpha+3)}{r_0^4}\Lambda+\frac{48\alpha^2(4\alpha^2-5\alpha+1)}{r_0^6}
, \label{xio}
\end{eqnarray}
In what follows, we discuss the properties of the solution (\ref{sol1}) in detail.
\subsection{Specific case: $\Lambda=0$} \label{tre}
\begin{figure}
\begin{center}
\includegraphics[scale=0.378]{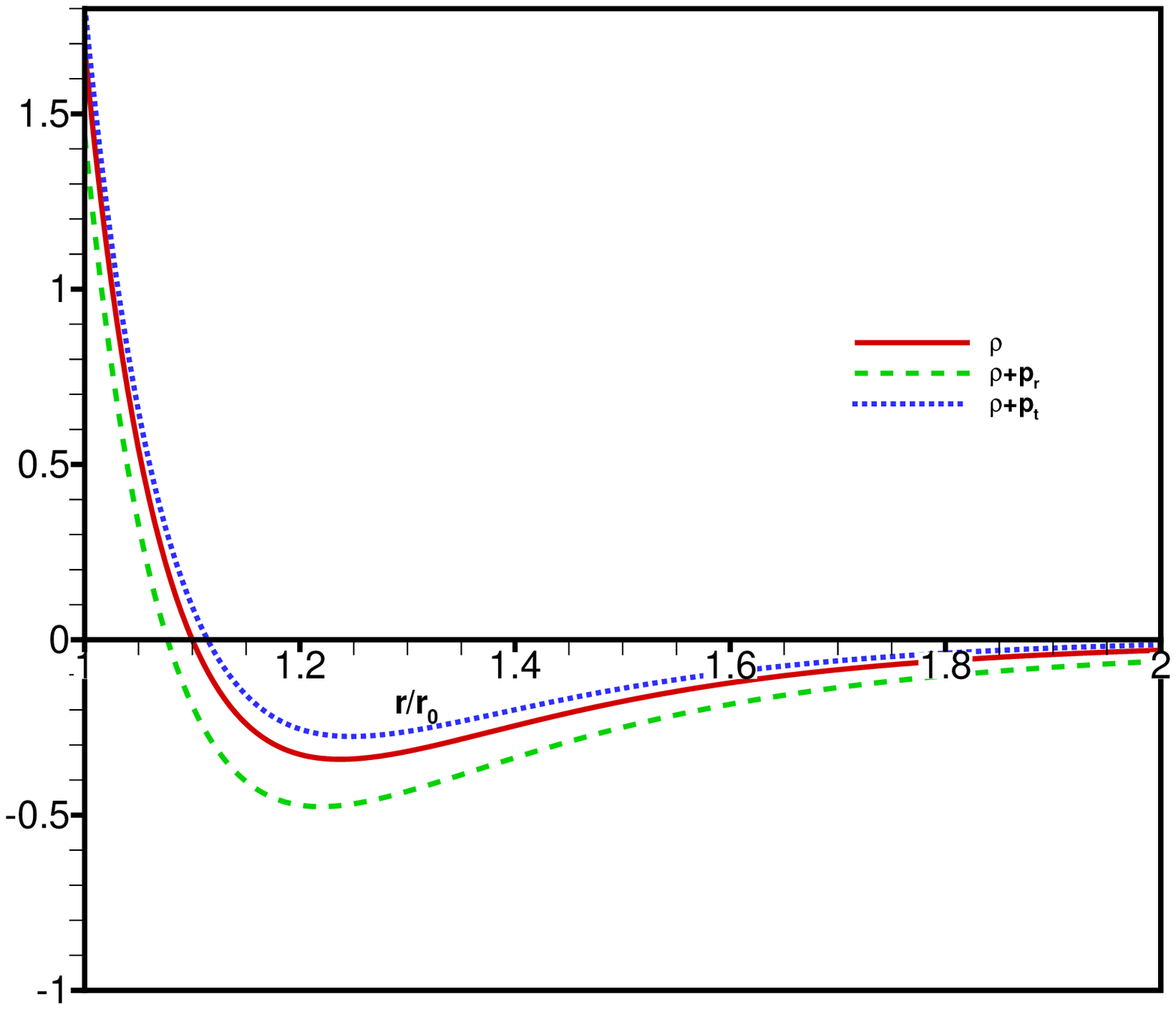}
\includegraphics[scale=0.378]{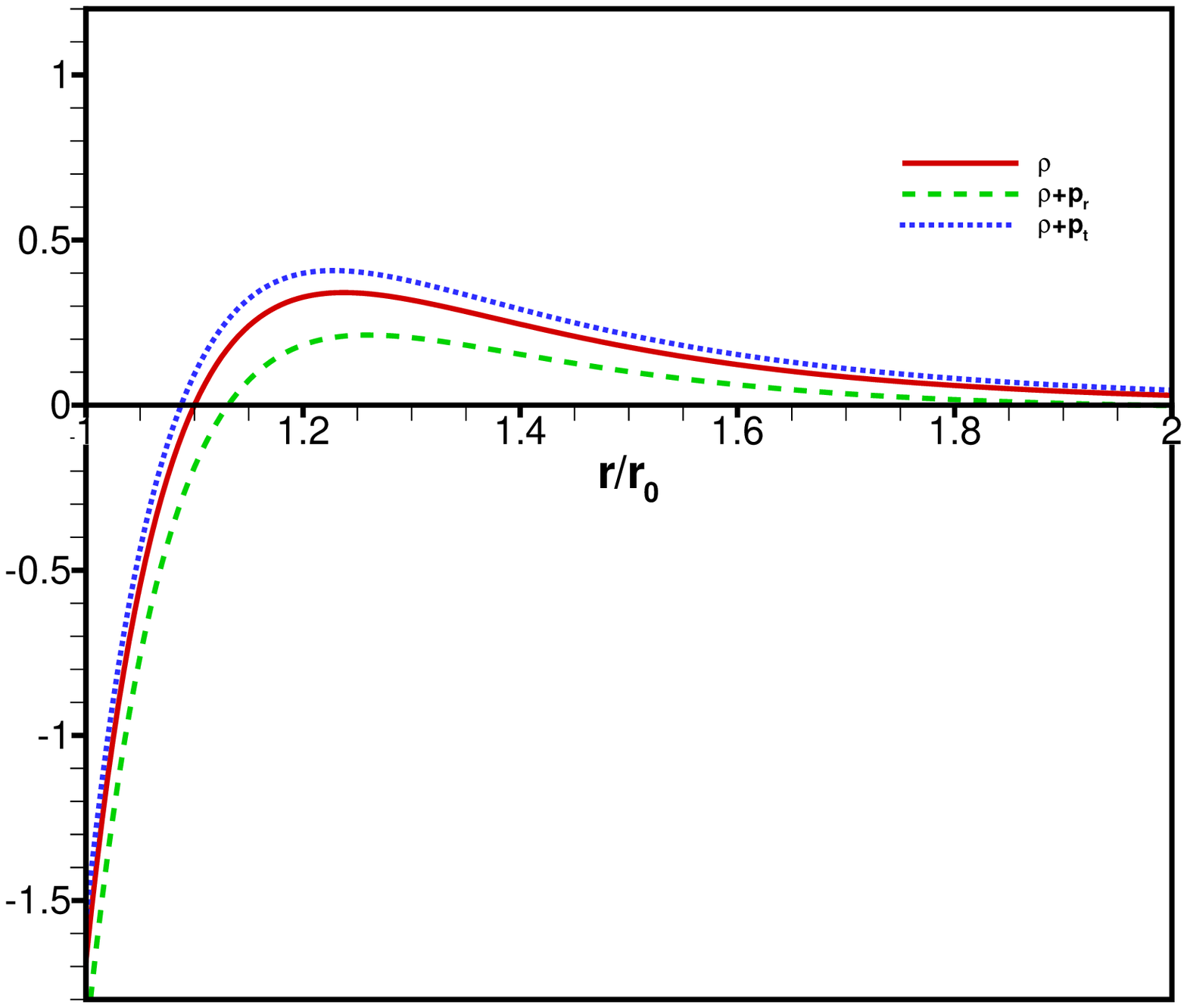}
\caption{The behavior of $\protect\rho $ (solid), $\protect\rho +p_{r}$ (dotted) and $\protect\rho +p_{t}$ (dashed) versus $r/r_{0}$. The model parameters are set as $\lambda=2$ (left panel), $\lambda=-2$ (right panel), $\alpha=-0.5$, $\Lambda=0$ and $r_0=2$. We choose units so that $\kappa=1$.}\label{wor2}
\end{center}
\end{figure}

Setting $\Lambda=0$ in Eq. (\ref{sol1}), we get the metric function as
\begin{align}
1-\frac{b(r)}{r}=1-\left(\frac{r}{r_0}\right)^{{\bf 2}\alpha}.
\end{align}
We note that condition $ b^{\prime}(r_0)<1$ in (\ref{bpr1}) implies that $\alpha<0$.  It is then clear that these solutions are asymptotically flat, i.e., ${b(r)}/{r}$ tends to zero as $r\rightarrow\infty$. For $\Lambda=0$ and in the limit of large values of $r$ coordinate, the $\xi_0$ coefficient in (\ref{rho11}) can be ignored in comparison to the last term and hence the energy density and pressure profiles for this case read
\begin{align}
\rho(r)=\frac{(1+2\alpha)r^{2(\alpha-1)}}{\kappa r_0^{2\alpha}},
\end{align}
\begin{align}
\rho(r)+p_r(r)=\frac{2\alpha r^{2(\alpha-1)}}{\kappa r_0^{2\alpha}},
\end{align}

\begin{align}
\rho(r)+p_t(r)=\frac{(\alpha+1)r^{2(\alpha-1)}}{\kappa r_0^{2\alpha}}.
\end{align}
This is effectively the {\sf GR} limit of {\sf ECG} which can be achieved in the limit of $\lambda\rightarrow0$. We therefore observe that for large values of $r$, the quantities $\rho(r)$ and $\rho+p_t$ are positive for $\alpha>-\frac{1}{2}$ and $\alpha>-1$, respectively. Hence, the tangential component of {\sf WEC} is satisfied for $-1/2<\alpha<0$. However the value of $\rho+p_r$ is always negative in the limit of large $r$ and thus, the radial component of {\sf WEC} is violated in this limit. Substituting $\Lambda=0$ into Eqs. (\ref{rho00})-(\ref{pt00}) we obtain, at the throat, $(4\alpha^2-5\alpha+1)>0$ for $\alpha<0$; thus the condition $\rho+p_r>0$
at the throat is respected for $\lambda >0$. Therefore, in order to satisfy the {\sf WEC} at the throat, we can choose suitable values of positive $\lambda$ such that $\rho$, $\rho+p_r$ and $\rho+p_t$ get positive values. It can also be seen that $\rho$, $\rho+p_r$ and $\rho+p_t$ have
real roots ($r_{i}$), where their values are positive in the interval $r_0<r<r_{i}$. The values of $r_i$ correspond to positive real roots of the following
equations
\bea\label{condswg}
\Sigma=0~~~~~~~~\textrm{for}~~\rho=0,\nn
2\Sigma-r_0^{4\alpha}r^4=0~~~~\textrm{for}~~~\rho+p_r=0,\nn
2\Sigma-\alpha r_0^{4\alpha} r^4=0~~~~~\textrm{for}~~~~\rho+p_t=0,
\eea 
where 
\begin{eqnarray}
\Sigma=24\kappa^2 \lambda \alpha (\alpha-1)(4\alpha-1)(5\alpha-3) r^{4\alpha}-24 \kappa^2 \lambda r_0^{2\alpha} \alpha (\alpha-1)(4\alpha-1)(4\alpha-3)r^{2\alpha} \notag + r_0^{4\alpha} (2\alpha+1)r^{4}.
\end{eqnarray}
In this case, we obtain $r_1=r_2=r_3=r_0 \left(\frac{4\alpha-3}{5\alpha-3}\right)^{1/2\alpha}$ when $\lambda$ is very large. The quantities $\rho$, $\rho+p_r$ and $\rho+p_t$ are sketched in Fig. (\ref{wor2}). Note that the components of {\sf EMT} tend to zero as $r$ tends to infinity. The left panel shows that for $\lambda>0$ the {\sf WEC} (and also {\sf NEC}) is violated in the vicinity of the wormhole throat, but for $\lambda <0$, it can be satisfied near the wormhole throat as it is shown in
the right panel. It is worth mentioning that for black hole solutions in {\sf ECG}, the coupling parameter $\lambda$ can affect the mass, entropy, Hawking temperature and horizon radius of the black hole, e.g., the horizon radius with $\lambda>0$ is greater than the Schwarzschild value~\cite{cub4}. In~\cite{observECG}, possible observational implications of {\sf ECG} have been explored and it is shown that, for spherically symmetric black holes, the corrections due to {\sf ECG} are remarkable on the scales near the horizon. However, the $\lambda$ parameter here affects only on the behavior of energy density (and consequently {\sf WEC}) and faraway from the wormhole configuration the {\sf ECG} corrections in energy density disappear. It is therefore reasonable to choose those values of $\lambda$ parameter for which the {\sf WEC} is satisfied, at least at the wormhole throat.

\subsection{General case: $\Lambda \neq 0$} 
Let us proceed with the general case in which $\Lambda\neq0$. Firstly, let us rewrite the metric function (\ref{sol1}) as 
\begin{align}
1-\frac{b(r)}{r}=1-\left(\frac{r}{r_0}\right)^{2\alpha}+\Lambda r_0^2 \left(\frac{r}{r_0}\right)^{2\alpha}-\Lambda r^2.
\end{align}
We note that this solution does not correspond to  an asymptotically flat spacetime, however, one may match this solution to an exterior vacuum geometry \cite{dml}. The flare-out condition (\ref{bpr1}) implies that we must have  $\Lambda> \frac{\alpha}{(\alpha-1)r_0^2}$ for $\alpha>1$ and  $\Lambda<\frac{\alpha}{(\alpha-1)r_0^2}$ for $\alpha<1$. In what follows, we analyze the physical properties and characteristics of these  wormholes along with checking the conditions under which the {\sf WEC} can be satisfied.
\par
\begin{figure}
\begin{center}
\includegraphics[scale=0.35]{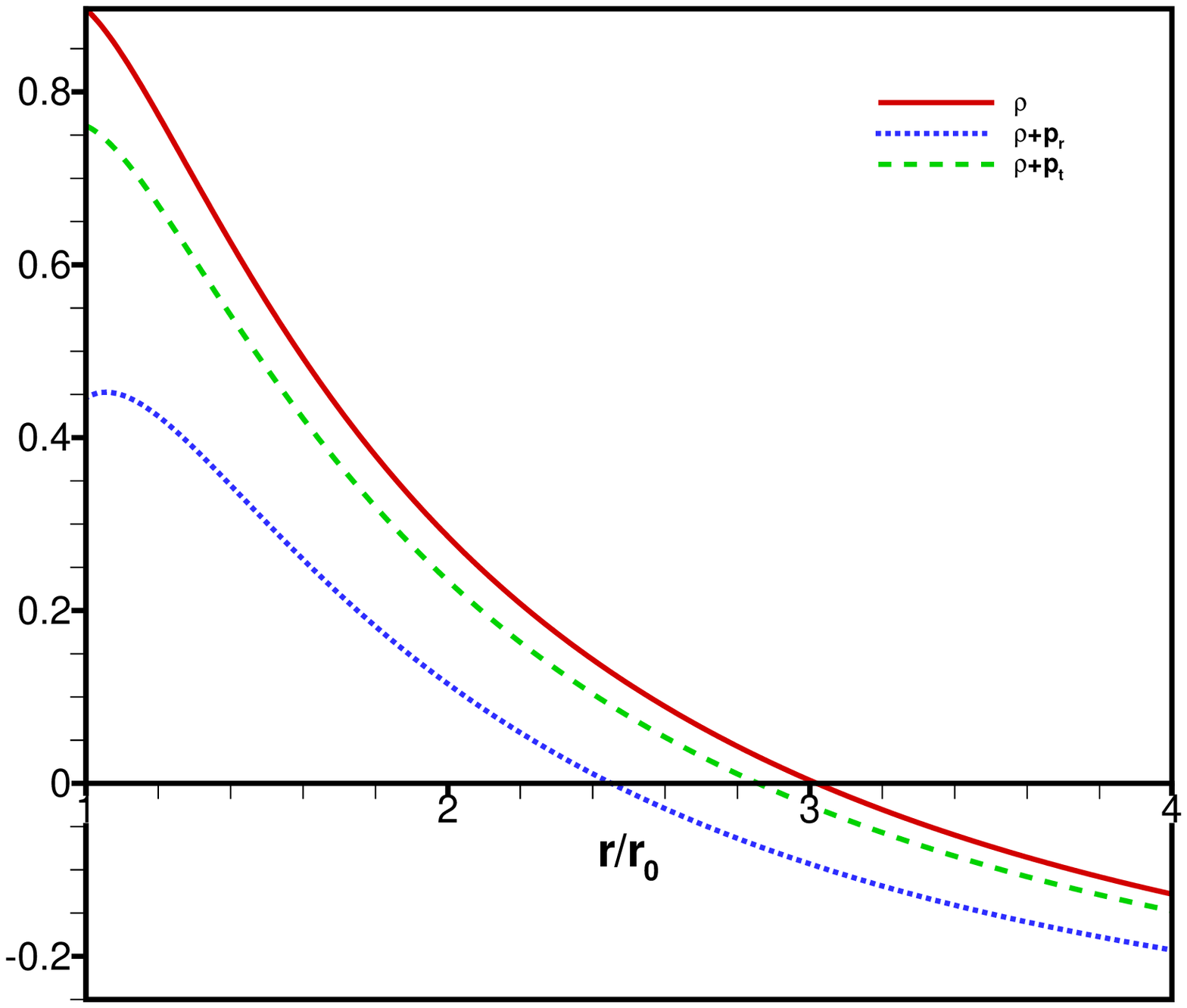}
\includegraphics[scale=0.35]{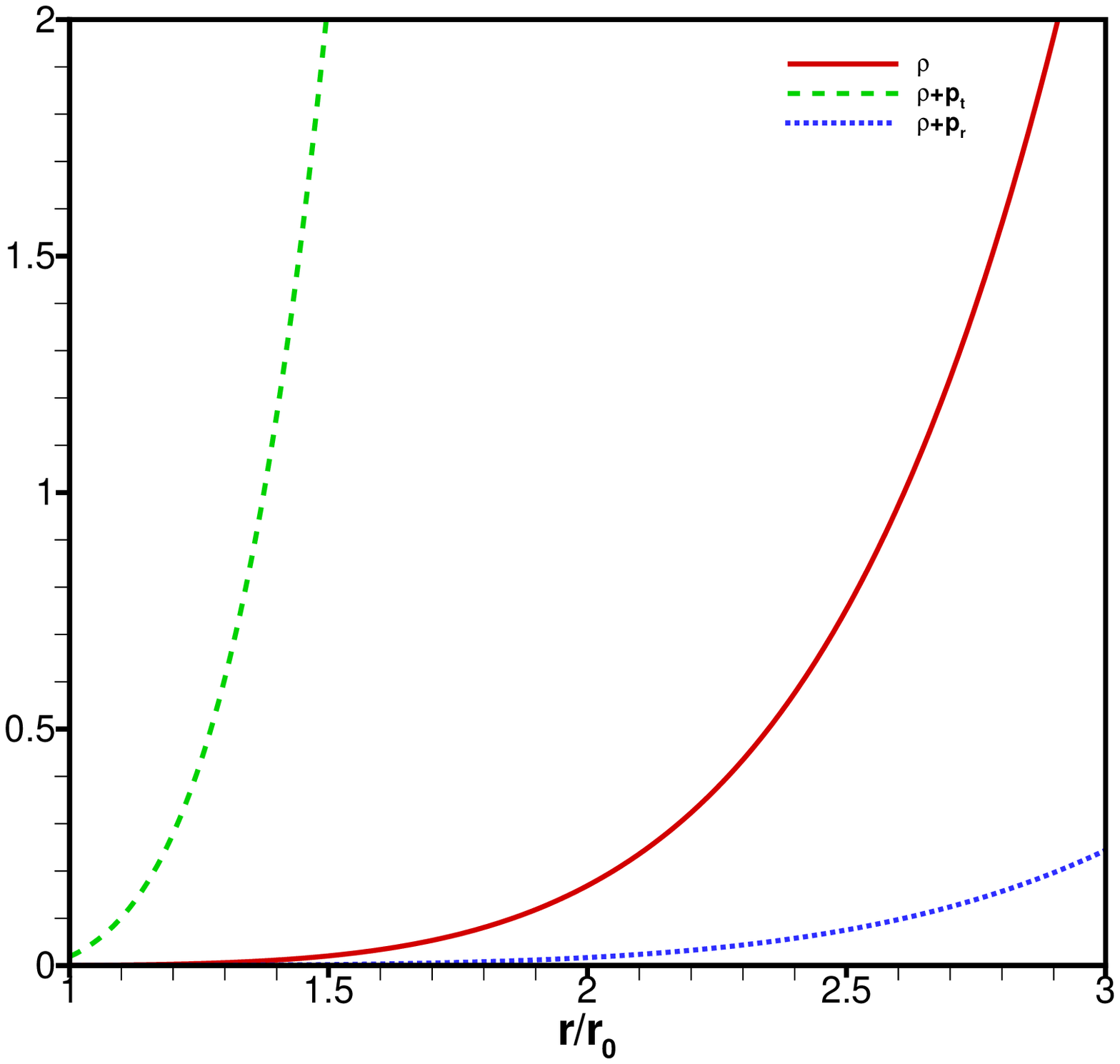}
\caption{The behavior of $\protect\rho $ (solid), $\protect\rho +p_{r}$ (dotted) and $\protect\rho +p_{t}$ (dashed) versus $r/r_{0}$ . The model parameters are set as $\lambda=-2$ (left panel), $\lambda=-0.2$ (right panel), $\alpha=0.3, 2$, $\Lambda=-0.2 , 1$ and $r_0=2$. We choose units so that $\kappa=1$.}\label{fig2ab}
\end{center}
\end{figure}
\begin{figure}
\hbox{\hspace{-1.2cm}\includegraphics[scale=0.45]{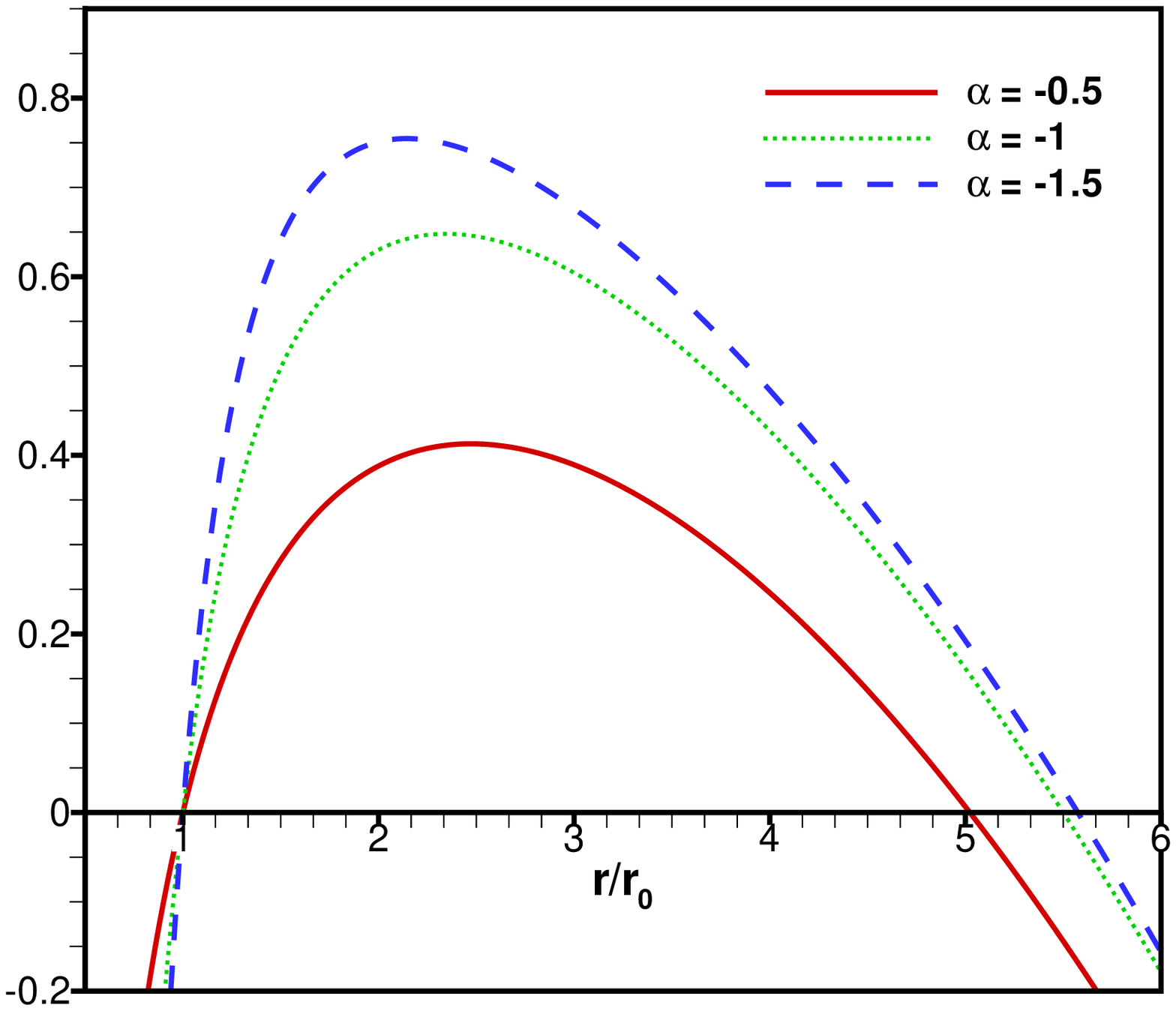}
\includegraphics[scale=0.45]{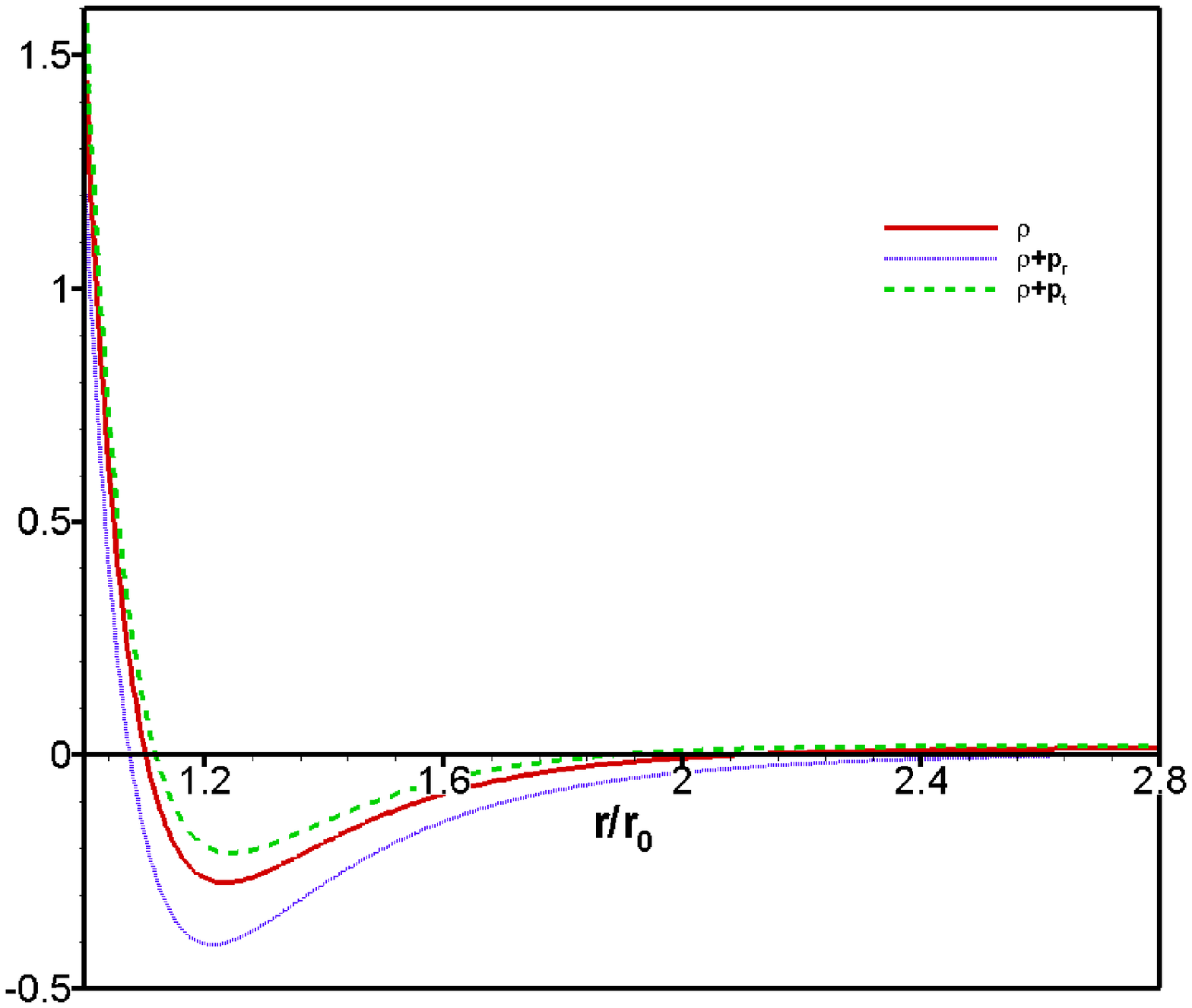}}
\caption{Left panel: The behavior of $1-b(r)/r$ versus $r/r_{0}$ for $\alpha=-0.5$ , $\alpha=-1$ , $\alpha=-1.5$ from down to up, respectively. Right panel: the behavior of $\protect\rho $ (solid), $\protect\rho +p_{r}$ (dotted) and $\protect\rho +p_{t}$ (dashed) versus $r/r_{0}$  for $\alpha=-0.5$. The model parameters are set as $\lambda=2$, $\alpha=-0.5$, $\Lambda=0.008$ and $r_0=2$. We choose the units so that $\kappa=1$.}\label{fig3ab}
\end{figure}

For $\alpha>1$, it is seen that for a suitable value of $\lambda<0$, the {\sf WEC} is satisfied throughout  the spacetime. For this case, the right panel of Fig. (\ref{fig2ab}) displays the behavior of $\rho$ , $\rho+p_r$ and $\rho+p_t$ that, as is seen, these quantities are positive throughout the spacetime, implying that the {\sf WEC} is satisfied for all values of $r$. In case we take $\alpha<0$ the value $\frac{\alpha}{(\alpha-1)r_0^2}$ is positive and therefore for $-\infty<\Lambda<\Lambda_c=\frac{\alpha}{(\alpha-1)r_0^2}$ the condition $b^{\prime}(r_0)<1$ is satisfied. To be a solution of a wormhole, the condition
$0<r-b(r)$ is also imposed. For the range of values $0<\Lambda<\Lambda_c$, condition $b(r)=r$ leads to two real and positive roots given by $r_{-}=r_0$ and $r_{+}=r_0{\left[\left(1+\frac{1}{\Lambda r_0^2-1}\right)\frac{1}{\alpha}\right]}^{\frac{1}{2(\alpha-1)}}$, and thus the spatial extension of this type of wormhole solution cannot be arbitrarily large. We then have a finite wormhole within the range
\begin{align}
r_{0}<r<r_0{\left[\left(1+\frac{1}{\Lambda r_0^2-1}\right)\frac{1}{\alpha}\right]}^{\frac{1}{2(\alpha-1)}}.
\end{align}
The left panel in Fig. (\ref{fig3ab}) shows that an increase in the value $\mid{\!\alpha\!}\mid$ enlarges the wormhole spatial extension. Note that for $-\infty<\Lambda<0$, we have Ads wormhole solution. In order to study energy conditions for these class of solutions we proceed with obtaining the behavior of quantities $\rho$, $\rho+p_r$ and $\rho+p_t$ at infinity. All of the quantities tend to the value $\Lambda$ as $r$ tends to infinity. Therefore, in the large limit of $r$, for $0<\Lambda<\Lambda_c$, the {\sf WEC} is satisfied. The right panel of Fig. (\ref{fig3ab}) shows that for $\lambda<0$ the {\sf WEC} (and also {\sf NEC}) is violated in the vicinity of wormhole throat, but for $\lambda>0$, these conditions can be satisfied near the throat.  In order to have normal matter at infinity, one can choose the positive value for $\Lambda$. We also note that the radius of throat can be determined by imposing energy conditions at the throat. To do so, one can use expressions (\ref{rho00})-(\ref{pt00}) to find three inequalities. Depending on the model parameters, these inequalities then decide the minimum and maximum values for $r_0$.

Finally, we can see that for $0<\alpha<1$ the term  $\frac{\alpha}{(\alpha-1)}$ is negative and therefor we have wormholes for negative value of $\Lambda$. We plot the quantities $\rho$ , $\rho+p_r$ and $\rho+p_t$ in the left panel of Fig. (\ref{fig2ab}) . All the quantities tend to $\Lambda_c=-0.2$ as $r$
tends to infinity. We have considered a suitable value for $\lambda<0$ in order to have normal matter distribution in the vicinity of the throat.

\section{Particle Trajectories Around the Wormhole}
\begin{figure}
	\begin{center}
		\includegraphics[scale=0.348]{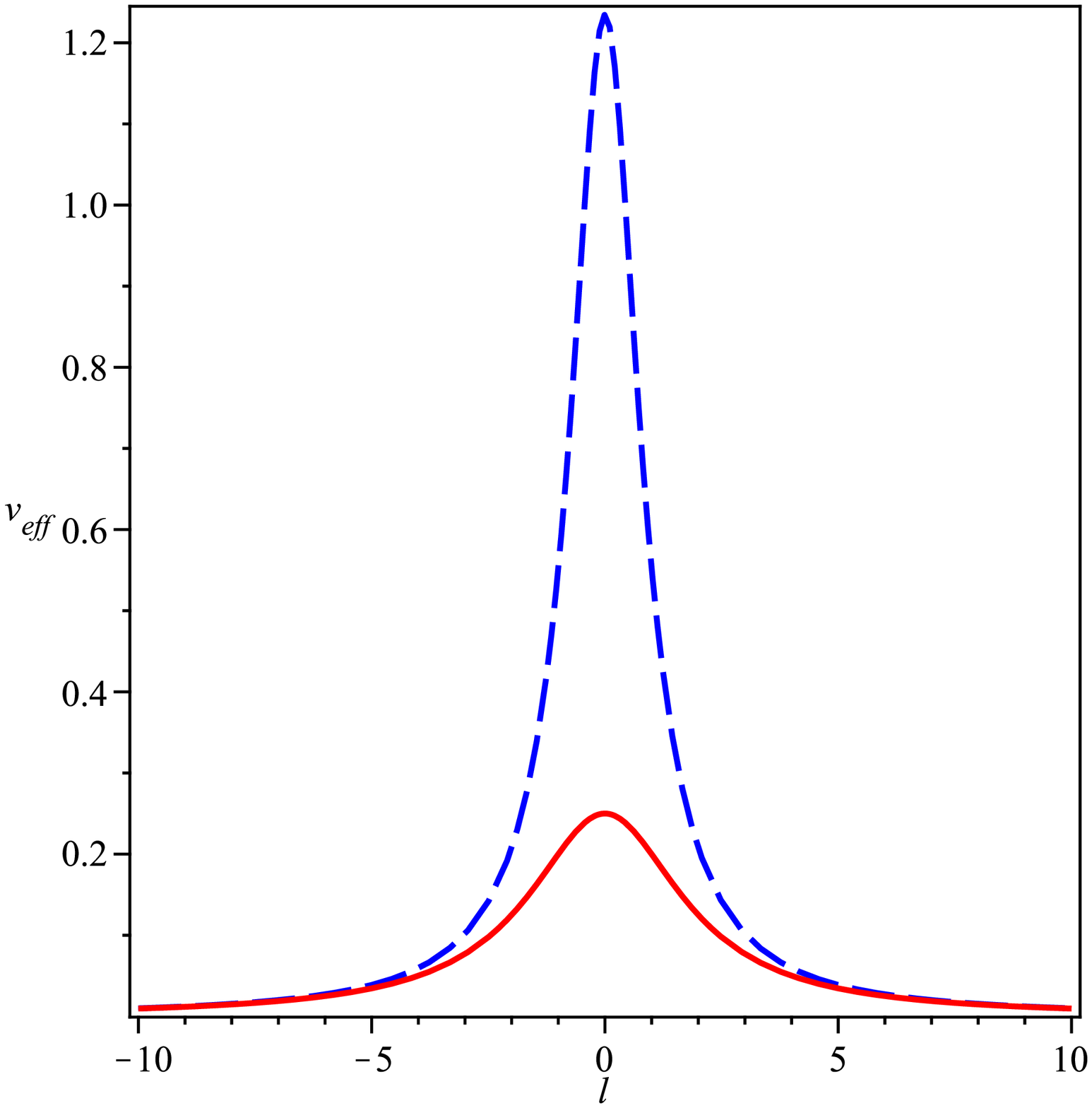}\label{woa}
		\includegraphics[scale=0.368]{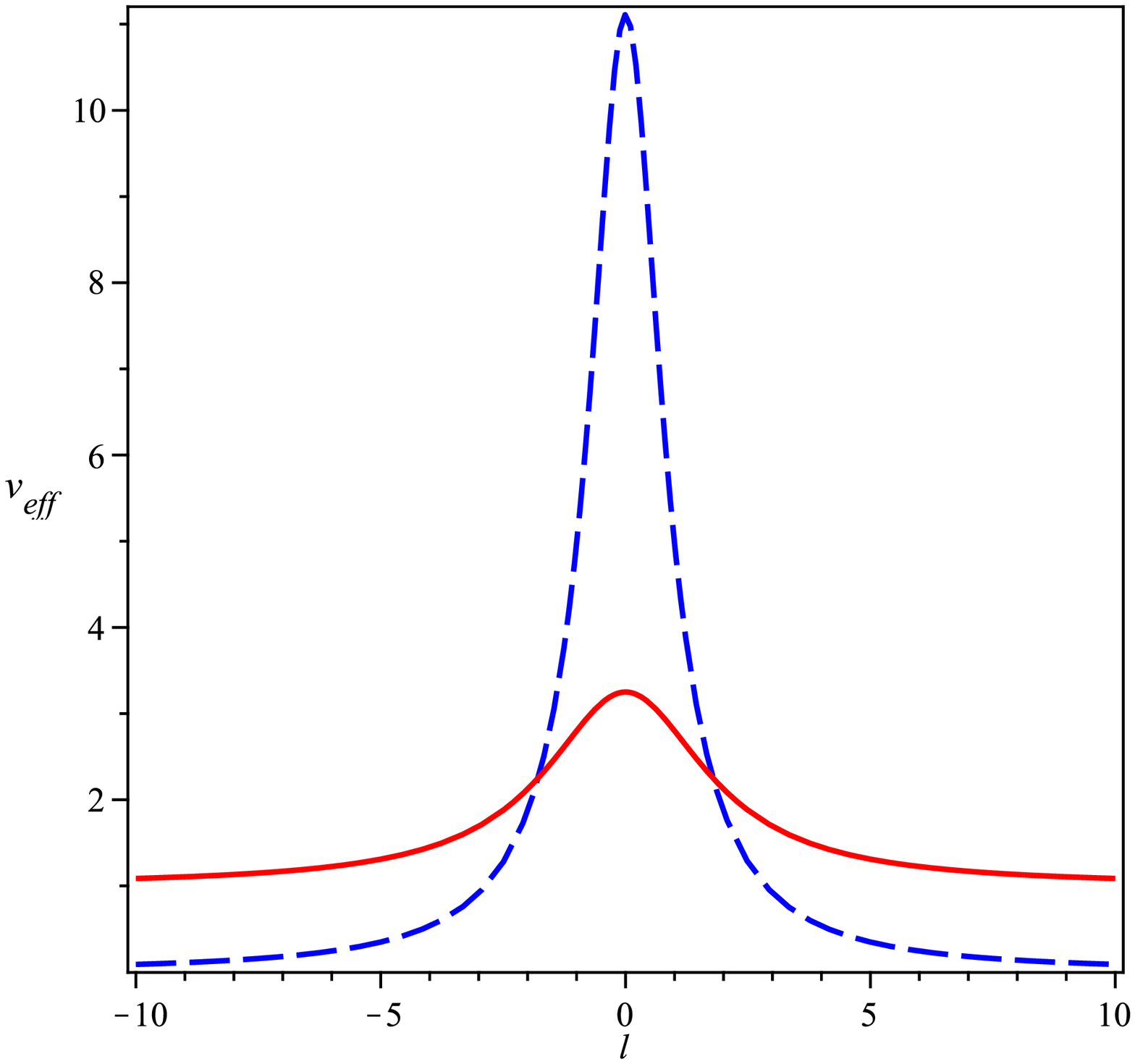}\label{wob}
		\caption{Effective potential $V_{{\rm eff}}$ for null (left panel) and timelike (righr panel) geodesics. In these plots we have the wormhole
			solutions with $L=3$ (the blue curve), $L=1$ (the red curve) and $\alpha=-1$.}\label{veffw}
	\end{center}
\end{figure}

\begin{figure}
	\begin{center}
		\includegraphics[scale=0.350]{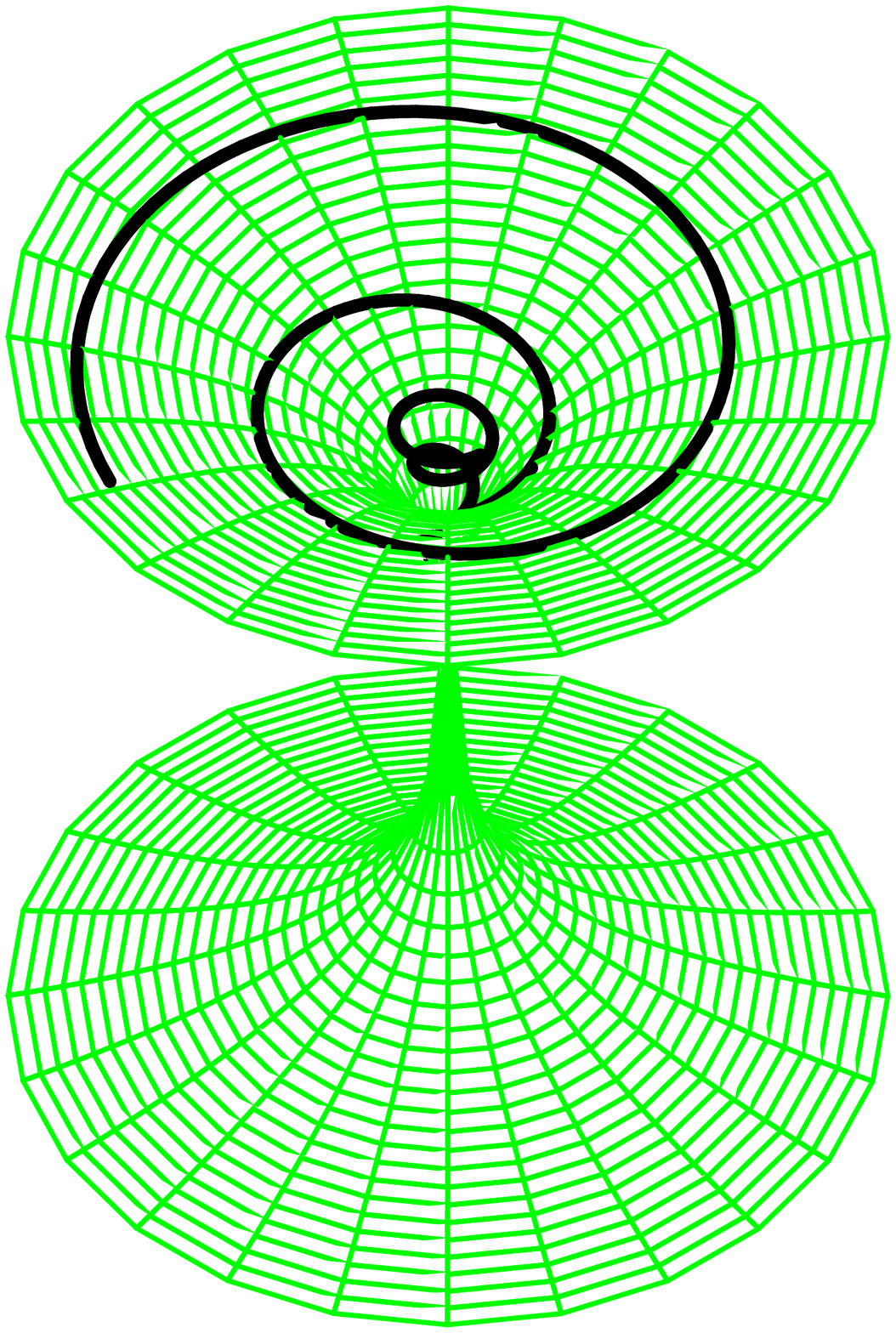}\label{wora}
		\includegraphics[scale=0.350]{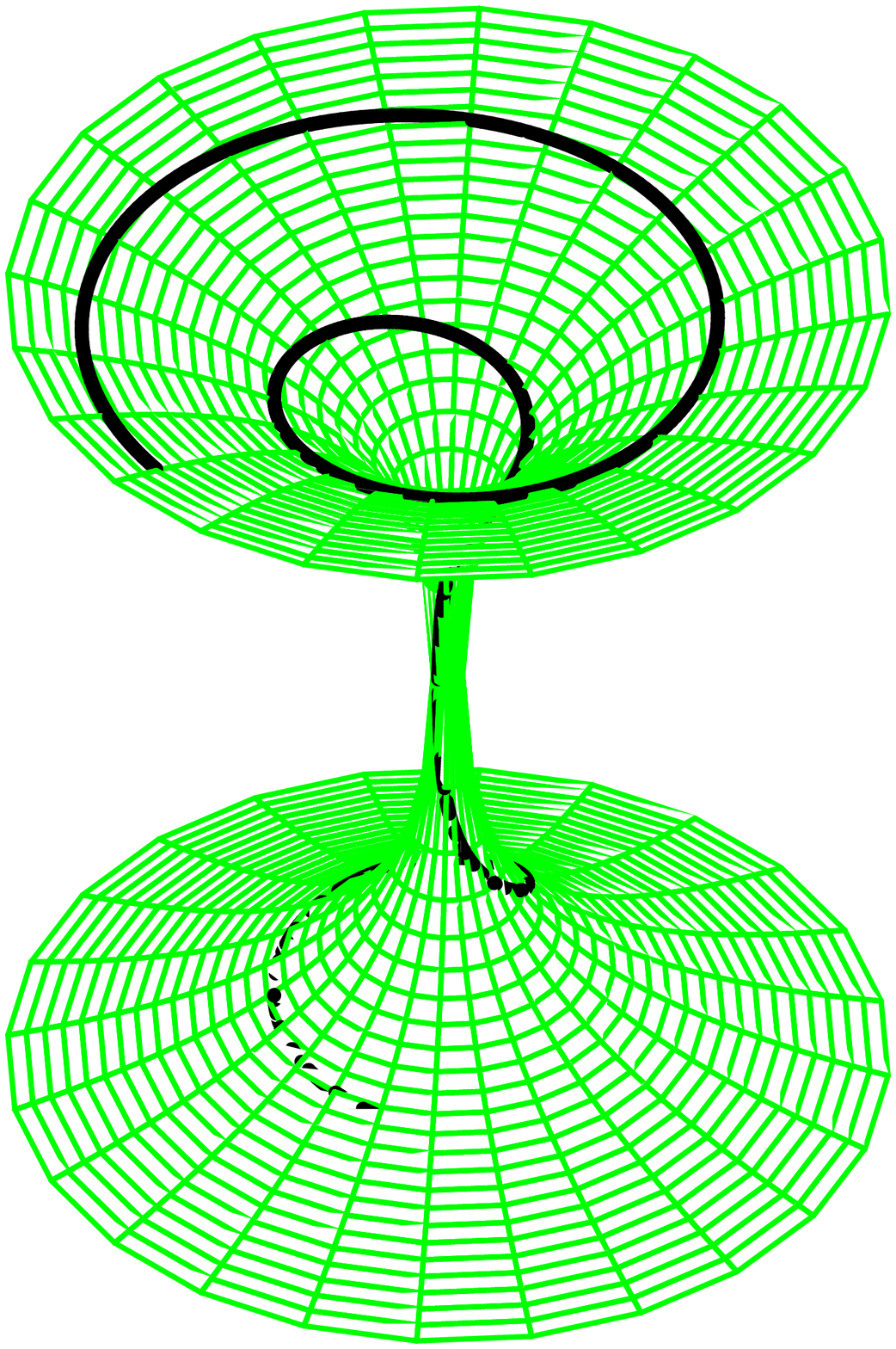}\label{worb}
		\caption{Embedding diagrams and behavior of null geodesic for $\alpha=-1$, $r_0=2$, $L=1$ with initial conditions $l_i=4$ and $\phi(0)=0$. The reflected and transmitted geodesics are shown in the left and right plots, respectively.}\label{geon}
	\end{center}
\end{figure}

\begin{figure}
	\begin{center}
		\includegraphics[scale=0.350]{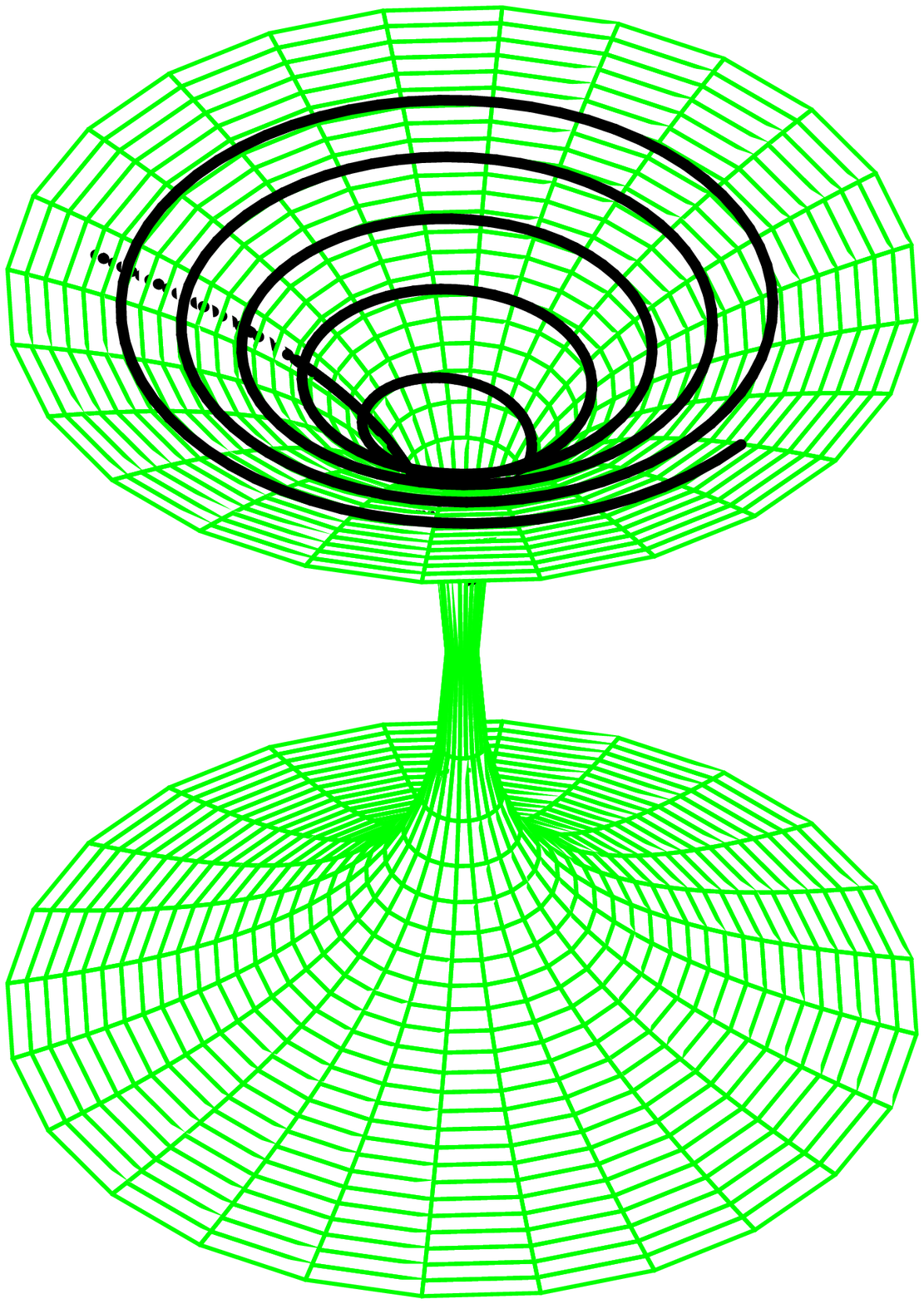}
		\includegraphics[scale=0.350]{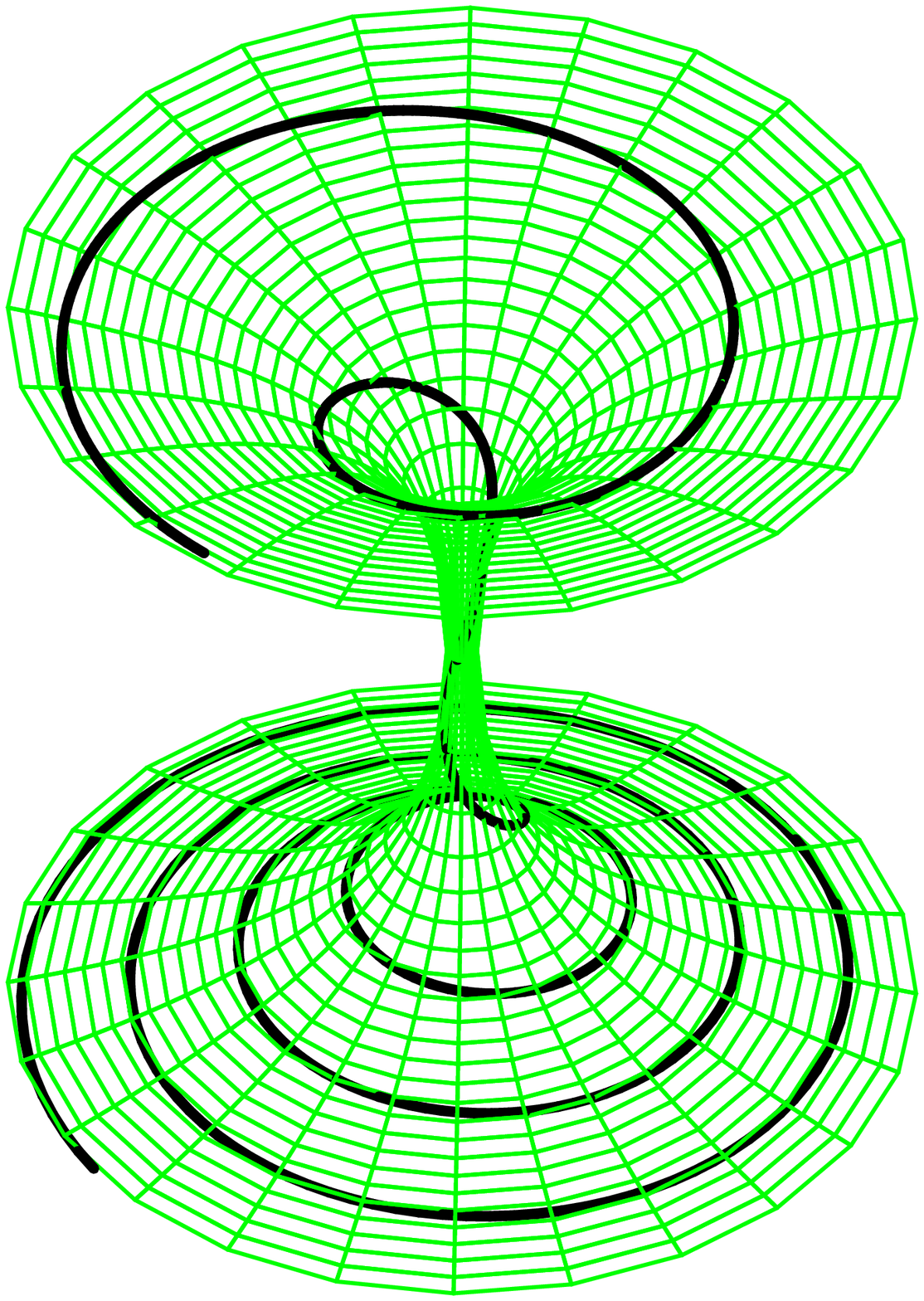}
		\caption{Embedding diagrams and behavior of timelike geodesics for $\alpha=-1$, $r_0=2$, $L=3$ with initial conditions $l_i=4$ and $\phi(0)=0$. The reflected and transmitted geodesics are shown in the left and right plots, respectively.}\label{geot}
	\end{center}
\end{figure}

In this section we analyze geodesic equations in {\sf ECG} wormhole spcetime described by the metric Eq. (\ref{sol1}), using  the Lagrangian formalism \cite{lagf}. Due to the spherical symmetry, it suffices to consider the equatorial plane $\theta=\pi/2$. The corresponding Lagrangian for metric (\ref{sol1}) is then found as
\begin{equation}
\mathfrak{ L} = g_{\mu\nu} \dot{x}^\mu \dot{x}^\nu= -\dot{t}^2+\frac{\dot{r}^2}{1-\frac{b(r)}{r}}+r^2\dot{\phi}^2   
\label{lag}
\end{equation}    
where a dot denotes derivative with respect to the affine parameter $\eta$. As the Lagrangian is constant along a geodesic we can consider $\mathfrak{L}(x^{\mu} , \dot{x}^\mu)={\epsilon}$ so that time-like and null geodesics correspond to ${\epsilon}=-1$ and ${\epsilon}=0$, respectively. Using the Euler-Lagrange equation, 
\begin{equation}\label{lag2}
\frac{d}{d\eta} \frac{\partial{\frak
		L}}{\partial\dot{x}^{\mu}}-\frac{\partial{\frak L}}{\partial
	x^{\mu}}=0,
\end{equation}
one can readily identify the following constants of motion
\begin{equation}\label{lag5}
-2\dot{t}=-2E \  \ {\rm and} \  \ 2r^2\dot{\phi}=2L,
\end{equation}
where $E$ is the energy and $L$ the angular momentum of the test particle.
Inserting these constants of motion into (\ref{lag}) we get
\begin{align}\label{radc}
\dot{r}^{2}&=\Big(1-\frac{b(r)}{r}\Big)\Big(E^{2}-\frac{L^{2}}{r^{2}}+\epsilon\Big).
\end{align}
It is convenient to rewrite Eq. (\ref{radc}) in terms of the proper radial distance
\begin{equation}\label{mt2}
l(r)=\pm\int_{r_0}^r\frac{dr}{(1-b(r)/r)^{1/2}},
\end{equation}
which is finite for all finite values of $r$. Note that the spacetime is extended in such a way that $l$ monotonically increases from
$-\infty$ to $+\infty$. $l<0$ or $l>0$ correspond to two parallel universes joined together via a throat at $l=0$. Using the proper radial distance, Eq. (\ref{radc}) takes the simple form 
\begin{align}
\label{eq:geodesics}
\dot{l}^{2}+V_{{\rm eff}}(L,l)=E^{2},
\end{align}
where the effective  potential is defined as
\begin{align}
\label{eq:potential}
V_{{\rm eff}}(L,l)=\frac{L^{2}}{r(l)^{2}}-\epsilon.
\end{align}
In what follows, we discuss the trajectory of particles around the wormhole, using the above form for the effective potential. In fact, geodesic equation (\ref{eq:geodesics}) can be interpreted as a classical scattering problem with a potential barrier $V_{{\rm eff}}(L,l)$. Moreover, using Eq. (\ref{lag5}) we can rewrite Eq. (\ref{eq:geodesics}) as an ordinary differential equation for orbital motion 
\begin{equation}\label{lag7}
\left(\frac{dl}{d\phi}\right)^2=\frac{{\dot{l}}^2}{\dot{{\phi}}^2}=\frac{{r(l)}^4}{L^2}\left[E^2- V_{{\rm eff}}(L,l)\right].
\end{equation} 
We note that, in traversable wormhole spacetimes, particles can travel through the throat of the wormhole from one asymptotically flat part of the manifold to other one. Then, a geodesic can pass through the throat into the other universe if 
\begin{align}
E^{2}>V_{{\rm eff}}(L,0)=\frac{L^{2}}{r_{0}^{2}}-\epsilon.
\end{align} 
Similarly, for a geodesic reflected back on the same universe by the potential barrier, we have $ E^{2}<V_{{\rm eff}}(L,0)$. In this case, there is a turning point at $l=l_{{\rm tu}}$ which is obtained by solving the following equation 
\begin{align}
E^{2}=V_{{\rm eff}}(L,l_{{\rm tu}}).
\end{align}
A generic feature of this effective potential is that it possesses a global maximum at the throat
\begin{align}
	\frac{dV_{{\rm eff}}}{dl}\Big|_{l=0}=0,\quad\quad \quad \frac{d^{2}V_{{\rm eff}}}{dl^{2}}\Big|_{l=0}=-\frac{L^{2}}{r^{4}}\Big(\frac{b(r)}{r}-b'(r)\Big)\Big|_{l=0}<0.
\end{align}
The flaring out condition leads to $\f{d^2V_{{\rm eff}}}{d\ell^2}<0$ at the throat. This clearly has an unstable orbit since it occurs at the maximum of
the potential for $E^{2}=V_{{\rm eff}}(L,l_{0})$. We note that these conditions are independent of whether the geodesic is null or timelike.
We now consider the wormhole solutions presented in section (\ref{tre}) and restrict ourselves to the class of wormholes with $\Lambda=0$. Setting the shape function $b(r)=r\left({r}/{r_0}\right)^{2\alpha}$ in Eq. (\ref{mt2}), we find
\begin{align}
\label{eq:embedding}
l(r)=\pm r\,\,{}_{2}F_{1}\big[\tfrac{1}{2},\tfrac{1}{2\alpha},1+\tfrac{1}{2\alpha},(r/r_{0})^{2\alpha}\big]-r_{0}\sqrt{\pi}\frac{\Gamma\left[\tfrac{2\alpha+1}{2\alpha}\right]}{\Gamma\left[\tfrac{1+\alpha}{2\alpha}\right]},
\end{align}
There is not an explicit representation for the effective potential $V_{{\rm eff}}(L,l)$ as a function of $l$, since Eq.~(\ref{eq:embedding}) cannot in general be solved for the radial coordinate as a function of proper distance. However, this issue can be numerically treated using standard techniques. Nevertheless, for $\alpha=-1$ we can obtain the behavior of effective potential as a function of the proper distance for timelike and null geodesics, see Fig.~(\ref{veffw}). We are interested in two kinds of trajectories for the obtained wormhole solutions. In the the first kind, test particles are not allowed to pass through the throat and remain within the original universe, whereas for the geodesics of the second kind the test particle passes through the throat from lower universe to the upper one. In order to visualize the trajectory of particles, we must solve numerically Eq.~(\ref{lag7}) with specified initial conditions. We note that the constants of motion $E$ and $L$ can be expressed by appropriate initial conditions. In Figs. (\ref{geon}) and (\ref{geot}) we have plotted trajectories of particles in the embedding diagram of the wormhole for timelike and null geodesics. In these plots, radius of the throat has been set as $r_0=2$ and the initial value for proper distance is fixed at $l_i=4$. In each graph, two kinds of trajectories for null Fig.~(\ref{geon}) and timelike Fig.~(\ref{geot}) geodesics are plotted with different initial conditions. We would also like to mention that although, for the present wormhole solutions, the effects of coupling parameter may not appear directly in the behavior of particle trajectories around the wormhole, these effects can be detected using the Raychaudhuri equation~\cite{Raychaudhurieq}, where the $\lambda$ parameter, by the virtue of energy density profile, would play its own role in the expansion rate of a congruence of timelike or lightlike geodesics. Not straightforwardly, but rather presenting additional information on the role of {\sf ECG} coupling parameter, more general wormhole solutions can be obtained considering nonzero redshift function. In this case, the effects of $\lambda$ parameter on particle trajectories can be better detected and the corresponding results can be compared with those of {\sf GR}. We hope to address these issues and other questions in future work.
\par
\begin{figure}
	\begin{center}
		\includegraphics[scale=0.380]{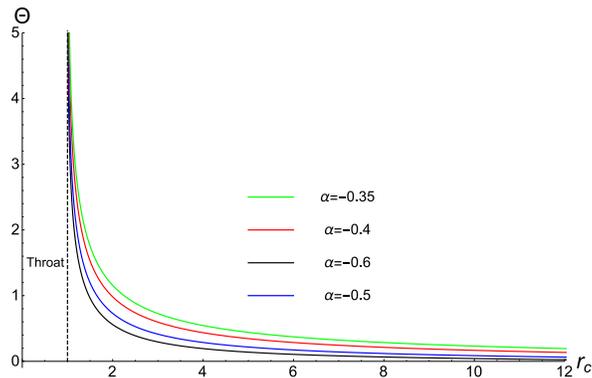}
		\caption{Deflection angle versus the closest distance approach for different values of {\sf EoS} parameter. The throat radius has been set as $r_0=1$.}\label{defanglefig}
	\end{center}
\end{figure}

A wormhole configuration has its own influences on its environment, an important aspect through which observational signatures of the wormhole can be surveyed. It is well known that the  gravitational lensing  is now one of useful tools to seek not only for dark and massive objects, but also wormholes. In this regard, we end this section by studying lensing effects of the obtained wormhole solutions. Let us consider a beam of light incoming from infinity, reaching the closest approach distance $r_{{\rm c}}$ from the center of the gravitating body and then emerges in another direction. Therefore, a light ray approaching the wormhole with the shape function $b(r)=r_0\left({r}/{r_0}\right)^{2\alpha+1}$ is scattered and bends with the deflection angle~\cite{Bozza}
\begin{equation}\label{defangleint}
\Theta(r_{{\rm c}},\alpha)=-\pi+2\int_{r_{\rm c}}^\infty \frac{r^{-(1+\alpha)}r_{{\rm c}}dr}{\sqrt{\left(r^{-2\alpha}-r_0^{-2\alpha}\right)\left(r^2-r_{{\rm c}}^2\right)}},
\end{equation}
where use has been made of Eq. (\ref{radc}) for null rays ($\epsilon=0$) along with the second part of Eq. (\ref{lag5}). The quantity $r_{{\rm c}}$ is defined as the distance at which $dr/d\phi=0$ and for the present model, it is related to the impact parameter $\beta=L/E$ as $r_{{\rm c}}=\beta$. Figure~\ref{defanglefig} shows how the light deflection angle depends upon the ratio of tangential to radial pressures, i.e., the {\sf EoS} parameter $\alpha$, and the distance of closest
approach (or correspondingly the impact parameter). We then observe that for $r_0<r_{{\rm c}}<\infty$ the quantity $\Theta(r_{{\rm c}},\alpha)$ is positive and finite. The deflection angle tends to zero as $r_{{\rm c}}\rightarrow\infty$, i.e., light ray is unaffected by the gravitating object. Moreover, the deflection angle increases as $r_{{\rm c}}\rightarrow r_0$ and diverges at the wormhole throat where an unstable photon sphere is present\footnote{As the impact parameter $\beta$ or equivalently the distance $r_{{\rm c}}$ are reduced, the deflection angle gets larger values. Further decreasing of the impact parameter makes the light ray to get extremely closer to the photon orbit causing the ray to wind up a large number of times before it emerges out. Consequently, as the closet distance approach reaches a critical value, the deflection angle diverges and the beam of light winds around a circular photon orbit indefinitely. In such a situation, circular photon orbits which collectively create a photon sphere, will satisfy $\dot{r}=\ddot{r}=0$~\cite{PerlickLRR}.}~\cite{grlensworm1}. Consequently, the wormhole structure may produce infinitely many relativistic images of an appropriately placed light source. This infinite sequence corresponds to infinitely many light rays whose limit curve asymptotically spirals towards the unstable photon sphere~\cite{HP2002}. Since the photon sphere is located at the throat, such a sphere may be detectable providing thus a setting to search for observational evidences of the wormhole. The integral (\ref{defangleint}) cannot be exactly solved for arbitrary values of {\sf EoS} parameter, however we can evaluate it in the weak field limit, i.e., far away enough from the gravitating body where $r_0\ll r_{{\rm c}}$. Introducing new variable $y=1-r_{{\rm c}}/r$ the integral (\ref{defangleint}) can be re-expressed in the following form
\begin{equation}\label{Thetay}
\Theta(\alpha)=-\pi+2\int_{0}^1dy\frac{\left(1-(1-y)^{-2\alpha}\zeta^{-2\alpha}\right)^{-\frac{1}{2}}}{\left(2y-y^2\right)^{\frac{1}{2}}},~~~~\zeta=\frac{r_0}{r_{{\rm c}}}.
\end{equation}
In the weak field regime where $\zeta\ll1$, we expand the numerator and perform the integration to get
\begin{eqnarray}\label{Thetayweakfl}
\Theta(\alpha)&=&2\alpha\sqrt{\pi}\zeta^{-2\alpha}\frac{\Gamma\left[-2\alpha\right]}{\Gamma\left[\frac{3}{2}-2\alpha\right]}\,{}_2 {F}_1\left[\frac{1}{2},1-2\alpha,\frac{3}{2}-2\alpha,-1\right]\nonumber\\&-&\frac{3}{4}\sqrt{\pi}\zeta^{-4\alpha}\frac{\Gamma\left[1-4\alpha\right]}{\Gamma\left[\frac{3}{2}-4\alpha\right]}\,{}_2 {F}_1\left[\frac{1}{2},1-4\alpha,\frac{3}{2}-4\alpha,-1\right]+{\mathcal O}\left(\zeta^6\right).
\end{eqnarray}
From the above solution we find that when $\alpha\rightarrow-1$, the deflection angle $\Theta(\alpha)\rightarrow\pi\zeta^2/4+9\pi\zeta^4/64$. Hence, in the limit where radial and tangential pressures cancel each other, the lensing characteristics of the wormhole structure mimic those of Ellis wormhole~\cite{grlensworm}.
 \section{Concluding Remarks}\label{concluding}
In the present study we constructed models of static wormholes with constant redshift function within the frame work of {\sf ECG} by considering ordinary matter distribution. Taking the {\sf EoS} of the matter fluid as a linear relation between radial and tangential pressures, we obtained exact traversable wormhole solutions which can be asymptotically flat, de-Sitter (dS) or AdS. For asymptotically flat solutions, the energy conditions in the vicinity of the throat can be satisfied by choosing suitable positive values of $\lambda$ parameter. Moreover, we showed that for $\lambda>0$, the {\sf WEC} can be satisfied near the throat for $r_0<r<r_3$ and a growth in the value of $\lambda$ parameter leads to increasing this range for radial coordinate. For dS solutions, the  energy conditions are fulfilled throughout the spacetime for $\alpha>1$ and $\lambda<0$. For AdS solutions, the energy conditions are satisfied for $0<\alpha<1$ and negative values of $\lambda$ parameter. Also, for $\lambda>0$ the energy conditions can be satisfied at the wormhole throat for both dS and AdS solutions provided that $\alpha<0$. It is noteworthy that asymptotically AdS wormhole solutions can be constructed in the context of holography. In~\cite{holoworm}, wormhole solutions with two asymptotic AdS boundaries are introduced and the effects of multi-trace deformations of the boundary quantum field theory on the wormhole state have been investigated. As discussed in~\cite{mt}, the imposition of causality and energy conditions prevents the Einstein-Rosen bridge from being traversable. However, as a result of the relationship between quantum information and quantum gravity in holography~\cite{qiqgholo} and particularly in the framework of Ads/CFT correspondence~\cite{adscft}, traversable wormhole configurations can be built in Ads/CFT by inserting a double-trace deformation on the boundary CFTs~\cite{doubletraceworm}. In this scenario quantum matter fields can provide the necessary negative energy (generated by explicit couplings between the two dual boundary CFTs) to retain the throat of the wormhole open and thus achieve its traversability. Nevertheless, traversable wormhole solutions presented here are obtained for classical matter fields that respect energy conditions for specific values of the model parameters. We also obtained embedding diagrams for wormhole solutions and the behavior of timelike and null geodesics in wormhole configuration were discussed. The study of wormhole geodesics is of great importance as it could help us to figure out possible observational effects that arise as a result of the scattering of particles on wormholes~\cite{SCATPARTWORM}. Likewise, such an investigation provides a promising way to test modified theories of gravitation as well as sufficient incentives for scientists towards probing wormhole structures in the universe. Work along this line has attracted many researchers to investigate observational aspects of a wormhole such as gravitational lensing~\cite{grlensworm,grlensworm1} and microlensing~\cite{microlens} effects, accretion disks around wormholes~\cite{accretionworm}, wormhole shadows~\cite{wormshadow} and several observables such as rotation curves~\cite{wormrotcurve}. 
\par
Traversable wormhole solutions in {\sf GR} can be obtained by considering some form of exotic fluid as the supporting matter for wormhole geometry. However, the situation could be completely different in the context of modified gravity theories where a modified {\sf EMT} could provide a suitable setting for traversable wormhole solutions, without the need of introducing any form of exotic matter. This objective was pursued within the framework of {\sf ECG} gravity and as we saw, depending on model parameters, a domain of existence of traversable wormhole solutions which respect energy conditions can be obtained. Beside the present model some other modified gravity theories have been investigated, whose geometrical attributes (not present in {\sf GR}) may provide a setting for investigating wormhole solutions without resorting to exotic matter. In this regard, much attempt has been made during the last decades to investigate the structure of wormholes in alternative theories of gravity such as, Einstein-Cartan theory~\cite{ECWORM}, modified gravity theories with curvature-matter coupling~\cite{WORMCMC} and multimetric gravity theories~\cite{WORMMMGT}. It is also worth mentioning that the solutions obtained in the present work can be generalized to the case of charged wormhole configuration. In this case, the Lagrangian for Maxwell field has to be added to the material part of the action (\ref{action}). Therefore, adding the extra source of matter (electromagnetic field) could enhance the degrees of freedom of the model (e.g., charge distribution may support the model against violation of {\sf WEC} and {\sf NEC} at the throat), providing then, a new class of traversable wormholes that respect energy conditions. Moreover, adding rotation to the present wormhole configurations may provide a different scenario in which, depending on the model parameters, wormhole structures with different properties (as compared to those of {\sf GR}) including their mass, angular momentum, quadrupole moment, and ergosphere can be constructed~\cite{FLoboBook}. However, dealing with the resultant field equations for the rotating case may not be an easy task and a more detailed study on this issue would be desirable.
\par
Finally, we would like to point out that, the stability or instability of celestial bodies is an important issue in gravitational physics. A traversable wormhole structure is of physical importance if it is stable against perturbations. In this regard, the stability of static wormholes has been surveyed utilizing specific equation of state or by considering a linearized radial perturbations around a static solution. Work along this line has been carried out for a thin-shell wormhole constructed from the Schwrazschild spacetime~\cite{PoiViss}. Stability of thin-shell wormhole in Einstein-Maxwell theory with a Gauss-Bonnet term has been studied in~\cite{wormemaxs} and stable and unstable static solutions of spherically symmetric thin shell wormholes supported by generalized Chaplygin gas has been investigated in~\cite{genchaps}, see also~\cite{FLoboBook,stab0} and references therein. An alternative technique for exploring the stability issue of wormhole structure is the thermodynamic stability. In this method, if the thin shell wormhole fulfills the first law of thermodynamics such that it respects the thermodynamic stability conditions in a successful way, and admits a meaningful heat capacity, then it can be considered as a stable wormhole configuration. Such an investigation has been performed for thin shell structures~\cite{thrmothinshell} and thin shell wormholes~\cite{thrmothinshellw}. In this regard, thermodynamic properties of the herein wormhole configuration could provide a testbed for thermodynamic stability of the  wormhole solutions, however, this study is beyond the scope of the present article and the results of future investigations will be reported as an independent work.


\begin{thebibliography}{99}
\bibitem{misner-wheeler} C. W. Misner and J. A. Wheeler, Ann. Phys. {\bf 2}, 525 (1957); \\C. W. Misner, Phys. Rev. {\bf 118}, 1110 (1959).
\bibitem{Wheelerworm} J. A. Wheeler, Ann. Phys. {\bf 2}, 604 (1957);\\ J. A. Wheeler, Geometrodynamics (Academic, New York, 1962).
\bibitem{mt} M. S. Morris and K. S. Thorne, Am. J. Phys. {\bf 56}, 395
(1988);\\ M. S. Morris, K. S. Thorne and U. Yurtsever, Phys. Rev. Lett. {\bf 61}, 1446 (1988).
\bibitem{khu} S. Kar, N. Dadhich, and M. Visser, Pramana J. Phys. {\bf 63}, 859
(2004); \\D. Hochberg and M. Visser, Phys. Rev. D {\bf 56}, 4745 (1997).
\bibitem{phantworm} F. S. N. Lobo, Phys. Rev. D {\bf 71}, 124022 (2005); Phys. Rev. D {\bf 71}, 084011 (2005); \\P. K. F. Kuhfittig, Class. Quant. Grav. {\bf 23}, 5853 (2006); \\S. Sushkov, Phys. Rev. D {\bf 71}, 043520 (2005).
\bibitem{phan-dark-sce} V. Sahni and A. A. Starobinsky, Int. J. Mod. Phys. D {\bf 9}, 373 (2000); \\S. M. Carroll, Living Rev. Rel. {\bf 4}, 1 (2001); \\P. J. E. Peebles and B. Ratra, Rev. Mod. Phys. {\bf 75}, 559 (2002);\\ V. Sahni, Class. Quantum Grav. {\bf 19}, 3435 (2002); \\T. Padmanabhan, Phys. Rep. {\bf 380}, 235 (2003); \\P. F. Gonzales-Diaz, Phys. Rev. D {\bf 65}, 104035 (2002).
\bibitem{phant1} R. R. Caldwell, Phys. Lett. B {\bf 545}, 23 (2002);\\ R. R. Caldwell, M. Kamionkowski, and N. N. Weinberg, Phys. Rev. Lett. {\bf 91}, 071301 (2003).
\bibitem{phant2} L. Amendola, Phys. Rev. Lett. {\bf 93}, 181102 (2004).
\bibitem{phant3} I. Brevik, S. Nojiri, S. D. Odintsov, and L. Vanzo, Phys.
Rev. D {\bf 70}, 043520 (2004); \\S. Nojiri and S. D. Odintsov,
Phys. Rev. D {\bf 70}, 103522 (2004).
\bibitem{thindynrot} N. M. Garcia, F. S. N. Lobo, M. Visser, Phys. Rev. D {\bf 86}, 044026 (2012);\\
M. G. Richarte, C. Simeone, Phys. Rev. D {\bf 76}, 087502 (2007);\\
E. Teo, Phys. Rev. D {\bf 58}, 024014 (1998);\\
P. E. Kashargin, S. V. Sushkov, Phys. Rev. D {\bf 78}, 064071 (2008);\\
S. Kar, D. Sahdev, Phys. Rev. D {\bf 53}, 722 (1996);\\
A. V. B. Arellano, F. S. N. Lobo, Class. Quantum Gravity {\bf 23}, 5811
(2006);\\
S. V. Sushkov, Y.-Z. Zhang, Phys. Rev. D {\bf 77}, 024042 (2008).
\bibitem{thi} E. Poisson and M. Visser, Phys. Rev. D {\bf 52}, 7318 (1995);\\ S. H. Mazharimousavi, M. Halilsoy, and Z. Amirabi, Phys. Rev. D {\bf 81}, 104002 (2010);\\ Class. Quantum Grav. {\bf 28}, 025004 (2011); \\M. R. Mehdizadeh, M. K. Zangeneh, and F. S. N. Lobo, Phys. Rev. D {\bf 92}, 044022 (2015).
\bibitem{glo1} F. S. N. Lobo, Phys. Rev. D {\bf 75}, 064027 (2007);\\ Classical
Quantum Gravity {\bf 25}, 175006 (2008);\\ F. S. N. Lobo and
M. A. Oliveira, Phys. Rev. D {\bf 80}, 104012 (2009);\\ N. M.
García and F. S. N. Lobo, Phys. Rev. D {\bf 82}, 104018 (2010);\\
N. M. García and F. S. N. Lobo, Classical Quantum Gravity
{\bf 28}, 085018 (2011);\\ C. G. Boehmer, T. Harko, and F. S. N.
Lobo, Phys. Rev. D {\bf 85}, 044033 (2012);\\ S. Capozziello, T.
Harko, T. S. Koivisto, F. S. N. Lobo, and G. J. Olmo, Phys.
Rev. D {\bf 86}, 127504 (2012);\\ A. G. Agnese and M. La Camera, Phys. Rev. D {\bf 51}, 2011
(1995);\\ K. K. Nandi, A. Islam, and J. Evans, Phys. Rev. D
{\bf 55}, 2497 (1997);\\ X. Yue and S. Gao, Phys. Lett. A {\bf 375},
2193 (2011);\\ F. S. N. Lobo and M. A. Oliveira, Phys. Rev. D
{\bf 81}, 067501 (2010);\\ S. V. Sushkov and S. M. Kozyrev, Phys.
Rev. D {\bf 84}, 124026 (2011).
\bibitem{mkl} M. R. Mehdizadeh, M. K. Zangeneh, and F. S. N. Lobo,
Phys. Rev. D {\bf 91}, 084004 (2015).
\bibitem{bd} A. G. Agnese and M. La Camera, Phys. Rev. D {\bf 51}, 2011 (1995);\\
K. K. Nandi, A. Islam, and J. Evans, Phys. Rev. D {\bf 55}, 2497 (1997); \\ F. S. N. Lobo and M. A. Oliveira, Phys. Rev. D {\bf 81}, 067501 (2010); \\S. V. Sushkov and S. M. Kozyrev, Phys. Rev. D {\bf 84}, 124026 (2011).
\bibitem{fr} F. S. N. Lobo and M. A. Oliveira, Phys. Rev. D {\bf 80}, 104012 (2009); \\N. M. Garcia and F. S. N. Lobo,
Phys. Rev. D {\bf 82}, 104018 (2010); \\N. Montelongo
Garcia and F. S. N. Lobo, Class. Quantum Grav. {\bf 28}, 085018 (2011);\\ M. Sharif and I. Nawazish, Ann. Phys. {\bf 389} 283 (2018);\\ J. B. Dent, D. A. Easson, T. W. Kephart and S. C. White, Int. J. Mod. Phys. D, {\bf 26} 1750117 (2017);\\ S. Bahamonde, M. Jamil, P. Pavlovic and M. Sossich, Phys. Rev. D {\bf 94}, 044041 (2016);\\ E. F. Eiroa and G. F. Aguirre, Phys. Rev. D {\bf 94}, 044016 (2016);\\ E. F. Eiroa and G. F. Aguirre, Eur. Phys. J. C {\bf 76} 132 (2016);\\ S. Bhattacharya and S. Chakraborty, Eur. Phys. J. C {\bf 77} 558 (2017);\\ M. Sharif and Z. Yousaf, Astrophys. Space Sci. {\bf 351} 351 (2014);\\ P. Pavlovic and M. Sossich, Eur. Phys. J. C {\bf 75}, 117 (2015).
\bibitem{bf} E. F. Eiroa and G. F. Aguirre, Eur. Phys. J. C {\bf 72}, 2240 (2012);\\ M. Richarte and C. Simeone, Phys. Rev. D {\bf 80}, 104033 (2009);\\ J. Y. Kim and M.-I. Park, {\bf 76}, 621 (2016);\\ R. Shaikh, Phys. Rev. D {\bf 92}, 024015 (2015);\\R. Shaikh, Phys. Rev. D {\bf 98}, 064033 (2018);\\ M. Azam, Astrophys. Space Sci. {\bf 361}, 96 (2016).
\bibitem{gmfl} M. R. Mehdizadeh, M. K. Zangeneh and F. S. N. Lobo, Phys. Rev. D {\bf 91}, 084004 (2015);\\ M. K. Zangeneh, F. S. N. Lobo, and M. H. Dehghani, Phys. Rev. D {\bf 92}, 124049 (2015);\\ T. Kokubu, H. Maeda and T. Harada, Class. Quant. Grav. {\bf 32}, 235021 (2015).
\bibitem{kl} V. D. Dzhunushaliev and D. Singleton, Phys. Rev. D {\bf 59}, 064018
(1999); \\J. P. de Leon, J. Cosmol. Astropart. Phys. {\bf 11},  013 (2009).
\bibitem{rastallworm} H. Moradpour, N. Sadeghnezhad and S. H. Hendi, Can. J. Phys. {\bf 95}, 1257 (2017).
\bibitem{kash} R. Shaikh and S. Kar, Phys. Rev. D {\bf 94}, 024011 (2016);\\X. Y. Chew, B. Kleihaus and J. Kunz , Phys. Rev. D {\bf 97}, 064026 (2018);\\S. Bahamonde, U. Camci, S. Capozziello and M. Jamil, Phys. Rev. D {\bf 94}, 084042 (2016);\\R. Shaikh and S. Kar, Phys. Rev. D {\bf 94}, 024011 (2016).
\bibitem{anac} A. Anabalon and A. Cisterna, Phys. Rev. D {\bf 85}, 084035 (2012);\\ J. P. S. Lemos, F. S. N. Lobo and S. Q. Oliveira, Phys. Rev. D {\bf 68}, 064004 (2003).
\bibitem{Wormnonc} F. Rahaman, A. Banerjee, M. Jamil, A. K. Yadav and H. Idris, Int. J. Theor. Phys., {\bf 53}, 1910 (2014);\\ M. Jamil, F. Rahaman, R. Myrzakulov, P. K. F. Kuhfittig, N. Ahmed and U. F. Mondal, J. Kor. Phys. Soc., {\bf 65}, 917 (2014);\\ M. Sharif and H. I. Fatima, Mod. Phys. Lett. A, {\bf 30} 1550142 (2015);\\ F. Rahaman, S. Ray, G. S. Khadekar, P. K. F. Kuhfittig, I. Karar, Int. J. Theor. Phys., {\bf 54}, 699 (2015);\\ M. Zubair, G. Mustafa, S. Waheed and G. Abbas, Eur. Phys. J. C, {\sf 77}, 680 (2017);\\ F. Rahaman, S. Karmakar, I. Karar and S. Ray, Phys. Lett. B, {\bf 746}, 73 (2015);\\ P. K. F. Kuhfittig, Int. J  Mod. Phys. D, {\bf 24}, 1550023 (2015);\\ M. Sharif and Kanwal Nazir, Mod. Phys. Lett. A {\bf 32}, 1750083 (2017);\\P. Bhara and F. Rahaman, Eur. Phys. J. C {\bf 74}, 3213 (2014).
\bibitem{modgrworm} M. Sharif and A. Ikram, Int. J. Mod. Phys. D {\bf 27}, 1750182 (2018);\\ P. H. R. S. Moraes and P. K. Sahoo, Phys. Rev. D {\bf 96}, 044038 (2017);\\P. K. Sahoo, P. H. R. S. Moraes and P. Sahoo, Eur. Phys. J. C {\bf 78}, 46 (2018);\\ M. Sharif and A. Ikram, Int. J. Mod. Phys. D {\bf 27}, 1750182 (2018);\\M. Sharif and K. Nazir, Ann. Phys. {\bf 393}, 145 (2018);\\ P. H. R. S. Moraesa, R. A. C. Correaa and R. V. Lobato, JCAP {\bf 07} 029 (2017);\\ H. Moradpour, N. Sadeghnezhad and S. H. Hendi, Can. J. Phys. {\bf 95}, 1257 (2017);\\ M. Zubair, Saira Waheed and Yasir Ahmad, Eur. Phys. J. C {\bf 76}, 444 (2016).
\bibitem{ste1} K. S. Stelle, Phys. Rev. D {\bf 16}, 953 (1977).
\bibitem{QFTCST} N. D. Birrell and P. C. W. Davies, "{\it Quantum Fields in Curved Space,}" (Cambridge University Press, Cambridge, England, 1982).
\bibitem{LEEAST} M. B. Greens, J. H. Schwarz and E. Witten, "{\it Superstring Theory,}" (Cambridge University Press, Cambridge, England, 1987);\\ D. Lust and S. Theusen, "{\it Lectures on String Theory,}" (Springer, Berlin, 1989);\\ J. Polchinski, "{\it String Theory,}" (Cambridge University Press, Cambridge, England,	1998).
\bibitem{LLOVE}  D. Lovelock, Aequationes mathematicae, {\bf 4}, 127 (1970);\\ D. Lovelock, J. Math. Phys. {\bf 12}, 498 (1971);\\ C. Charmousis, Lect. Notes Phys. {\bf 769}, 299 (2009);\\ T. Padmanabhan and D. Kothawala,
Phys. Rept. {\bf 531}, 115 (2013).
\bibitem{GBTERM} B. Zwiebach, Phys. Lett. B {\bf 156}, 315 (1985);\\ B. Zumino, Phys. Rep. {\bf 137}, 109 (1986);\\D. G. Boulware and S. Deser, Phys. Rev. Lett., {\bf 55}, 2656 (1985);\\ J. T. Wheller, Nucl. Phys.
B {\bf 268}, 737 (1986).
\bibitem{hol1} J. M. Maldacena, Int. J. Theor. Phys. {\bf 38}, 1113 (1999);\\ E. Witten, Adv. Theor. Math. Phys. {\bf 2}, 253 (1998).
\bibitem{cos1}  S. Nojiri and S. D. Odintsov, Phys. Rept. {\bf 505}, 59 (2011);\\ T. P. Sotiriou and V. Faraoni, Rev. Mod. Phys. {\bf 82}, 451
(2010);\\ T. Clifton, P. G. Ferreira, A. Padilla and C. Skordis, Phys. Rept. {\bf 513}, 1 (2012).
\bibitem{cfteqs} R. C. Myers and A. Sinha, JHEP {\bf 01}, 125 (2011);\\ P. Bueno and R. C. Myers, JHEP {\bf 08}, 068 (2015);\\  H. Liu and A. A. Tseytlin, Nucl. Phys. B {\bf 533}, 88 (1998);\\ A. Buchel, J. Escobedo, R. C. Myers, M. F. Paulos, A. Sinha, and M. Smolkin, JHEP {\bf 03}, 111 (2010).
\bibitem{qttheory} R. C. Myers and B. Robinson, JHEP {\bf 08}, 067 (2010).
\bibitem{flovel} P. Bueno, P. A. Cano, A. O. Lasso, and P. F. Ramirez,
JHEP {\bf 04}, 028 (2016);\\ A. Karasu, E. Kenar, and
B. Tekin, Phys. Rev. D {\bf 93}, 084040 (2016).
\bibitem{con1} P. Bueno and P. A. Cano, Phys. Rev. D {\bf 94}, 104005 (2016).
\bibitem{ghostfreecubic} P. Bueno, P. A. Cano, V. S. Min and M. R. Visser, Phys. Rev. D {\bf 95}, 044010 (2017).
\bibitem{cub4} P. Bueno and P. A. Cano, Phys. Rev. D {\bf 94}, 124051 (2016).
\bibitem{rub1} R. A. Hennigar and R. B. Mann, Phys. Rev. D {\bf 95}, 064055 (2017).
	
\bibitem{high2} P. Bueno and P. A. Cano, Class. Quant. Grav. {\bf 34}, 175008 (2017);\\ J. Ahmed, R. A. Hennigar, R. B. Mann and M. Mir, JHEP {\bf 05}, 134 (2017);\\ P. Bueno and P. A. Cano, Phys. Rev. D {\bf 96}, 024034
(2017).
\bibitem{linea1}S. Sushkov, Phys. Rev. D {\bf 71}, 043520 (2005);\\ F. S. N. Lobo, Phys. Rev. D {\bf 71}, 084011 (2005); {\bf 73}, 064028 (2006); {\bf 75}, 024023 (2007);\\ A. De Benedictis, R. Garattini, and F. S. N.
Lobo, Phys. Rev. D {\bf 78}, 104003 (2008);\\ F. S. N. Lobo, F.
Parsaei, and N. Riazi, Phys. Rev. D {\bf 87}, 084030 (2013).
\bibitem{trc1}  S. Kar and D. Sahdev, Phys. Rev. D {\bf 52}, 2030 (1995).
\bibitem{anch1} L. A. Anchordoqui, S. E. Perez Bergliaffa, and D. F. Torres, Phys. Rev. D {\bf 55}, 5226 (1997).
\bibitem{frh} F. Rahaman, M. Kalam, M. Sarker, A. Ghosh and B. Raychaudhuri, Gen. Rel. Grav. {\bf 39}, 145 (2007);\\  N. M. Garcia and F. S. N. Lobo, Phys. Rev. D {\bf 82}, 104018 (2010);\\ P. H. R. S. Moraes and P. K. Sahoo, Phys. Rev. D {\bf 96}, 044038 (2017).
\bibitem{observECG} R. A. Hennigar, M. B. Jahani Poshteh, R. B. Mann, Phys. Rev. D {\bf 97}, 064041 (2018).
\bibitem{dml} M. S. R. Delgaty and R. B. Mann, Int. J. Mod. Phys. D {\bf 04}, 231 (1995);\\ F. S. N. Lobo, Class. Quant. Grav. {\bf 21},
4811 (2004).
\bibitem{lagf} W. Rindler, Relativity, Special, General and Cosmology (Oxford University Press, New York, 2001).
\bibitem{Raychaudhurieq} S. Kar,  Phys. Rev. D {\bf 52}, 2036 (1995);\\N. J. Poplawski, Phys. Lett. B {\bf 687} 110 (2010);\\ L. H. Ford, T. A. Roman, Phys. Rev. D {\bf 87}, 085001 (2013);\\ B. Shoshany, arXiv:1907.04178 [gr-qc].
\bibitem{Bozza} V. Bozza, Phys. Rev. D {\bf 66}, 103001 (2002).
\bibitem{PerlickLRR} V. Perlick, Living Rev. Relativity, {\bf 7}, 9 (2004).
\bibitem{grlensworm1} R. Shaikh, P. Banerjee, S. Paul and T. Sarkar, Phys. Lett. B {\bf 789} 270 (2019).
\bibitem{HP2002} W. Hasse and V. Perlick, Gen. Relativ. Gravit. {\bf 34}, 415 (2002).
\bibitem{grlensworm} N. Tsukamoto, T. Harada, and K. Yajima, Phys. Rev. D {\bf 86}, 104062 (2012);
\bibitem{holoworm} P. Betzios, E. Kiritsis and O. Papadoulaki, JHEP, 06 (2019) 042.
\bibitem{qiqgholo} G.'t Hooft, \lq{}\lq{}{\it Dimensional reduction in quantum gravity,}\rq{}\rq{} arXiv:gr-qc/9310026 [gr-qc];\\
L. Susskind, J. Math. Phys. {\bf 36}, 6377 (1995);\\R. Bousso, Rev. Mod. Phys., {\bf 74} 825 (2002).
\bibitem{adscft} J. M. Maldacena, Adv. Theor. Math. Phys, {\bf 2}, 231 (1998);\\
S. S. Gubser, I. R. Klebanov, and A. M. Polyakov, Phys. Lett. B {\bf 428}, 105 (1998);\\
E. Witten, Adv. Theor. Math. Phys. {\bf 2}, 253 (1998);\\
O. Aharony, S. S. Gubser, J. Maldacena, H. Ooguri, and Y. Oz, Phys. Rept. {\bf 323}, 183 (2000).
\bibitem{doubletraceworm} P. Gao, D. L. Jafferis and A. C. Wall, JHEP, (2017)2017: 151;\\ J. Maldacena, D. Stanford, and Z. Yang, Fortsch. Phys. {\bf 65}, 1700034 (2017);\\ J. Maldacena and X.-L. Qi,  arXiv:1804.00491 [hep-th].
\bibitem{SCATPARTWORM} A. Kirillov and E. Savelova, Universe {\bf 4}, 35 (2018).
\bibitem{microlens} M. Safonova, D. F. Torres, and G. E. Romero, Phys. Rev. D {\bf 65}, 023001 (2001);\\ F. Abe, Astrophys. J., {\bf 725}, 787 (2010);\\ N. Tsukamoto and T. Harada, Phys. Rev. D {\bf 95}, 024030 (2017).
\bibitem{accretionworm} T. Harko, Z. Kovacs, and F. Lobo, Phys. Rev. D {\bf 78}, 084005 (2008);\\ C. Bambi, Phys. Rev. D {\bf 87}, 084039 (2013).
\bibitem{wormshadow} H. Falcke, F. Melia, and E. Agol, Astrophys. J. {\bf 528}, L13 (2000);\\P. G. Nedkova, V. K. Tinchev, and S. S. Yazadjiev, Phys. Rev. D {\bf 88}, 124019 (2013);\\T. Ohgami and N. Sakai, Phys. Rev. D {\bf 91}, 124020 (2015); \\A. Abdujabbarov, B. Juraev, B. Ahmedov and Z. Stuchlik, Astrophys. Space Sci. {\bf 361}, 226 (2016); \\R. Shaikh, Phys. Rev. D {\bf 98}, 024044 (2018); \\P. V. P. Cunha and C. A. R. Herdeiro, Gen. Relativ. Gravit. {\bf 50}, 42 (2018). 
\bibitem{wormrotcurve} V. Bozza and A. Postiglione, JCAP {\bf 1506}, 036 (2015).
\bibitem{ECWORM} K. A. Bronnikov and A. M. Galiakhmetov, Grav. Cosmol., {\bf 21}, 283 (2015); Phys. Rev. D {\bf 94}, 124006 (2016);\\ M. R. Mehdizadeh and A. H. Ziaie, Phys. Rev. D {\bf 96}, 124017 (2017); Phys. Rev. D {\bf 95}, 064049 (2017);\\ E. Battista, E. Di Grezia, M. Manfredonia and G. Miele, Eur. Phys. J. Plus {\bf 132}, 537 (2017).
\bibitem{WORMCMC} T. Harko, F. S. N. Lobo, M. K. Mak and S. V. Sushkov, Phys. Rev. D {\bf 87}, 067504 (2013);\\ Z. Yousaf, M. Ilyas, and M. Z.-ul-Haq Bhatti, Eur. Phys. J. Plus {\bf 132}, 268 (2017);\\ P. H. R. S. Moraes and P. K. Sahoo, Phys. Rev. D {\bf 97}, 024007 (2018);\\ P. K. Sahoo, P. H. R. S. Moraes and P. Sahoo, Eur. Phys. J. C {\bf 78}, 46 (2018);\\ Z. Yousaf, M. Ilyas and M. Z. Bhatti, Mod. Phys. Lett. A, {\bf 32}, 1750163 (2017).
\bibitem{WORMMMGT} M. Hohmann, Phys. Rev. D {\bf 89}, 087503 (2014).
\bibitem{FLoboBook} F. S. N. Lobo (Editor), \lq{}\lq{}{\it Wormholes, Warp Drives and Energy Conditions,}\rq{}\rq{} Springer (2017).
\bibitem{PoiViss} M. Visser Phys. Rev. D {\bf 39}, (1989) 3182;\\ E. Poisson and M. Visser. Phys. Rev. D {\bf 52}, 7318 (1995).
\bibitem{wormemaxs} M. Thibeault, C. Simeone and E. F. Eiroa, Gen. Rel. Gravit. {\bf 38}, 1593 (2006).
\bibitem{genchaps} E. F. Eiroa, Phys. Rev. D {\bf 80}, 044033 (2009).
\bibitem{stab0} M. Sharif, M. Azam, J. Phys. Soc. Jpn., {\bf 81}, 124006 (2012);\\M. Sharif, S. Mumtaz, Astrophys. Space Sci., {\bf 352}, 729 (2014);\\ M. Sharif, M. Azam, Eur. Phys. J. C {\bf 73}, 2407 (2013).
\bibitem{thrmothinshell} J. P. S. Lemos and G. M. Quinta, Phys. Rev. D {\bf 88}, 067501 (2013);\\
J. P. S. Lemos, F. J. Lopes, M. Minamitsuji and J. V. Rocha, Phys. Rev. D {\bf 92}, 064012 (2015);\\
J. P. S. Lemos, G. M. Quinta, and O. B. Zaslavskii, Phys. Rev. D {\bf 91}, 104027 (2015);\\
J. P. S. Lemos, O.B. Zaslavskii, Phys. Lett. B {\bf 695}, 37 (2011);\\
J. P. S. Lemos, G. M. Quinta, O. B. Zaslavskii, Phys. Lett. B {\bf 750}, 306 (2015).
\bibitem{thrmothinshellw} S. D. Forghani, S. H. Mazharimousavi and M. Halilsoy, arXiv:1812.04340 [gr-qc].
\end{thebibliography}
\end{document}